\documentclass[10pt]{article}

\usepackage{jheppub}
\usepackage{graphicx,epsfig,subfigure}
\usepackage{epigraph}
\usepackage{float}
\usepackage[usenames]{color} 
\usepackage{amssymb} 
\usepackage{amsmath} 
\usepackage{bbold}
\usepackage{booktabs}
\usepackage{slashed}
\usepackage[utf8]{inputenc} 
\usepackage{multirow}
\usepackage[table]{xcolor}

\newcommand{\bmtx}{\begin{pmatrix}}
\newcommand{\emtx}{\end{pmatrix}}
\newcommand{\be}{\begin{eqnarray}}
\newcommand{\ee}{\end{eqnarray}}
\newcommand{\bea}{\begin{eqnarray}\begin{aligned}}
\newcommand{\eea}{\end{aligned}\end{eqnarray}}
\newcommand{\tr}{\text{Tr}}

\newcommand{\gev}{\text{GeV}}
\newcommand{\tev}{\text{TeV}}

\newcommand{\mev}{\text{MeV}}

\newcommand{\vckm}{V_{\text{CKM}}}

\newcommand{\tH}{\widetilde{H}}

\newcommand{\pth}{\ensuremath{p_{T}^{h}}}

\newcommand{\mcO}{\ensuremath{\mathcal{O}}}

\newcommand{\diag}{\text{diag}}

\usepackage{tikz}
\usepackage[compat=1.1.0]{tikz-feynman}
\usetikzlibrary{positioning}
\usetikzlibrary{decorations.text}
\usetikzlibrary{decorations.pathmorphing}
\usetikzlibrary{calc}
\usetikzlibrary{shapes.misc}
\tikzset{
        >=latex,
    photon/.style={decorate, decoration={snake}, draw=black, thick},
    fermionnoarrow/.style={draw=black, postaction={decorate}, thick},
    scalar/.style={draw=black, postaction={decorate}, decoration={markings,mark=at position .55 with {\arrow{>}}}, thick, dashed},
    scalarnoarrow/.style={draw=black, postaction={decorate},  thick, dashed},
    fermion/.style={draw=black, postaction={decorate},decoration={markings,mark=at position .55 with {\arrow{>}}}, thick},
    gluon/.style={decorate, draw=black, decoration={coil,amplitude=4pt, segment length=5pt}, thick},
    vertex/.style={draw,shape=circle,fill=black,minimum size=3pt,inner sep=0pt},
    fillvertex/.style={draw,shape=circle,fill=black,minimum size=6pt,inner sep=0pt},
    openvertex/.style={draw,shape=circle,minimum size=5pt,inner sep=0pt},
    blob/.style={draw=gray,shape=circle,fill=black,minimum size=8pt,inner sep=0pt},
    redvertex/.style={draw=red,shape=circle,fill=red,minimum size=3pt,inner sep=0pt},
    cross/.style={cross out, draw=black,thick, minimum size=5pt, inner sep=0pt, outer sep=0pt}
}
\usepackage{feynmp-auto}

\newcommand{\lamfc}{\lambda_{\rm FC}}
\newcommand{\invlamfc}{\tilde{\lambda}_{\rm FC}}
\newcommand{\BR}{{\rm BR}}

\title{
Aligned Yet Large Dipoles: a SMEFT Study
}

\author[a,b,c,d]{Quentin Bonnefoy,}
\emailAdd{qbonnefoy@unistra.fr}

\author[d,e]{Jonathan Kley,}
\emailAdd{jonathan.kley@desy.de}

\author[f,d]{Di Liu,}
\emailAdd{liudisy@gmail.com}

\author[g]{Alejo N. Rossia,}
\emailAdd{alejo.rossia@manchester.ac.uk}

\author[d,h]{Chang-Yuan Yao}
\emailAdd{chang.yuan.yao@desy.de}

\affiliation[a]{Universit\'e de Strasbourg, CNRS, IPHC UMR7178, 23 rue du Loess, 67037 Strasbourg, France}
\affiliation[b]{Berkeley Center for Theoretical Physics, Department of Physics, University of California, Berkeley, CA 94720, USA}
\affiliation[c]{Theoretical Physics Group, Lawrence Berkeley National Laboratory, Berkeley, CA 94720, USA}
\affiliation[d]{Deutsches Elektronen-Synchrotron DESY, Notkestr. 85, 22607 Hamburg, Germany}
\affiliation[e]{Institut für Physik, Humboldt-Universität zu Berlin, 12489 Berlin, Germany}
\affiliation[f]{Laboratoire d'Annecy-le-Vieux de Physique Th\'eorique, CNRS -- USMB, BP 110 Annecy-le-Vieux, F-74941 Annecy, France}
\affiliation[g]{Dept. of Physics and Astronomy, University of Manchester, Oxford Road, Manchester M13 9PL, U.K.}
\affiliation[h]{School of Physics, Nankai University, Tianjin 300071, China}

\abstract{
We study a non-universal flavor scenario at the level of the Standard Model Effective Field Theory, according to which the matrix of Wilson coefficients $c_{uW}$ of an up-type electroweak quark dipole operator is aligned with the up-type Yukawa coupling. Such an alignment usually follows from the assumption of Minimal Flavor Violation (MFV), away from which we step by allowing the entries of $c_{uW}$ to be sizable along the first quark generations. A particular example, which we refer to as ``inverse hierarchy MFV", features Wilson coefficients inversely proportional to quark masses, and arises from BSM models respecting MFV and containing heavy fields that replicate the mass hierarchy of SM quarks. We then analyze the phenomenology driven by $c_{uW}$ at colliders and at lower-energy flavor experiments. 
We show that precision measurements of the process $pp\rightarrow W h\rightarrow \gamma\gamma\ell\nu$ at FCC-$hh$ could set an upper bound on $|c_{uW}|\lesssim\mathcal{O}(10^{-2})(\Lambda/\tev)^{2}$, with $\Lambda$ the cutoff of the effective field theory. This bound is an order of magnitude stronger than the existing LHC bounds.
Moreover, we estimate that $W h\rightarrow b\bar b \ell\nu$ at HL-LHC could also give competitive bounds.
In the low-energy regime, we consider bounds arising from rare kaon decays, which turn out to be loose, $|c_{uW}^{11}|<\mathcal{O}(1)(\Lambda/\tev)^{2}$.
We finally demonstrate that our flavor and operator assumptions can be derived from a weakly-coupled UV model, which we choose to simultaneously illustrate the UV origin of inverse hierarchy MFV.
}

\begin{document}

\begin{flushright}
DESY-24-033\\
HU-EP-24/09\\
LAPTH-011/24\\
COMETA-2024-004
\end{flushright}
\maketitle
\flushbottom

\section{Introduction}

New high-energy physics (NP), which is expected for several compelling, theoretical and observational reasons, could manifest itself in a variety of ways. Under the assumption that it couples significantly to the Standard Model (SM) degrees of freedom, two main scenarios have been considered: the new particles are light enough to be produced at colliders, or they are too heavy. In the latter case, NP models can be matched onto effective field theories (EFTs), which can often be taken to be the Standard Model Effective Field Theory (SMEFT)~\cite{Buchmuller:1985jz,Grzadkowski:2010es,Murphy:2020rsh,Li:2020gnx,Li:2020xlh,Harlander:2023psl,Liao:2016hru,Liao:2020jmn} (see~\cite{Alonso:2015fsp,Falkowski:2019tft,Cohen:2020xca} for exceptions), whose cutoff corresponds to the scale of NP. 

Taken at face value, the SMEFT has, already at dimension-6 order,
a great number of free parameters. Although one may consider all parts of this parameter space as equally probable 
up to theoretical constraints on its physical region (see {\it e.g.}, ~\cite{Arzt:1994gp,Einhorn:2013kja,Craig:2019wmo,Adams:2006sv,Zhang:2018shp,Remmen:2019cyz,Zhang:2020jyn}), the properties of the SM and many experimental measurements suggest that NP has a highly non-trivial structure. This is especially true in the flavor sector, where the existence of three generations of matter, together with the very hierarchical quark masses and mixings, suggest that a mechanism is at play at higher energies. Different foundational paradigms have been proposed in that direction, which require for instance new family-dependent ``horizontal" broken symmetries~\cite{Froggatt:1978nt,Leurer:1992wg,Leurer:1993gy}, extra dimensions of space~\cite{Arkani-Hamed:1999ylh,Grossman:1999ra,Dvali:2000ha,Gherghetta:2000qt,Huber:2000ie} or SM couplings to a strong sector that generate strongly hierarchical fermion kinetic terms~\cite{Nelson:2000sn,Davidson:2007si}. All those constructions predict very strong hierarchies in the flavored coefficients of the SMEFT~\cite{Isidori:2009px,Efrati:2015eaa,Aharony:2010ch,Csaki:2009wc,Bordone:2019uzc,Faroughy:2020ina,Greljo:2022cah} which, for the physicist interested in producing the associated new particles in experiments, is very fortunate: were the flavored SMEFT coefficients random, they would impose bounds on the scale of NP (or, to be precise, on the scale of flavor violation) much beyond the reach of foreseeable colliders~\cite{Gabbiani:1996hi,UTfit:2007eik,Isidori:2010kg,Crivellin:2013hpa,Pruna:2014asa,Feruglio:2015yua,Silvestrini:2018dos,Aebischer:2020dsw,Ardu:2021koz}. This is mostly due to the natural suppression of flavor-changing neutral currents (FCNC) in the SM~\cite{Glashow:1970gm}, whose compatibility with observations sets strong bounds on the NP scale for arbitrary flavored SMEFT coefficients. This conclusion is avoided in the aforementioned NP scenarios, which, in IR model-independent language, set the flavored SMEFT coefficients to values that make FCNC compatible with a cutoff at a couple of TeVs. One can also contemplate structures which prevent large FCNC in NP models that do not address the origin of quark masses and mixing hierarchies. For instance, symmetries have been invoked to avoid strong FCNC constraints in two-Higgs doublet models (2HDM)~\cite{Glashow:1976nt}. 

The same suppression of FCNC can be obtained without reference to a specific NP scenario, upon assuming the existence of suitable building blocks in the flavor sector appropriately distributed over the flavored SMEFT coefficients. The most prominent assumption is that of Minimal Flavor Violation (MFV)~\cite{Chivukula:1987py,Hall:1990ac,Buras:2000dm,DAmbrosio:2002vsn,Isidori:2010kg}, where it is assumed that only the SM Yukawa couplings break the $U(3)^5$ flavor symmetry between all generations of matter. This implies a Cabibbo-Kobayashi-Maskawa (CKM) and Yukawa suppression of flavor-violating processes (see~\cite{Feldmann:2008ja,Kagan:2009bn} for refinements accounting for the large top Yukawa), and captures for instance the low-energy effects of flavor-blind NP. Less stringent assumptions have also been explored, for instance that of an approximate $U(2)^3$ symmetry of the light quark generations~\cite{Agashe:2005hk,Barbieri:2012uh,Isidori:2012ts}, on which the strongest flavor bounds apply. $U(1)^9$ symmetries have also been discussed~\cite{Antaramian:1992ya,Hall:1993ca}, while extensive studies of flavor symmetries and spurions in the SMEFT can be found in~\cite{Faroughy:2020ina,Degrande:2021zpv,Kobayashi:2021pav,Greljo:2022cah}. 

When the minimal number of spurions is considered, all those flavor assumptions generate flavored SMEFT coefficients somewhat aligned with those of the SM Yukawas, {\it i.e.}, reproducing in part the mass and mixing hierarchies. However, this is not the only way of mitigating the strength of flavor constraints on the NP scale. In this work, we are interested in flavor scenarios with very suppressed FCNC, whose predictions nevertheless deviate significantly from the aforementioned scenarios, in particular from MFV. In addition to FCNC, MFV suppresses chirality flips of the light generations by their small Yukawas. Therefore, even without noticeable FCNC, violation of that scaling represents an unambiguous signal that NP is not flavor blind; this is what we are after in this work. 

A natural flavor setting which realizes the above is found in aligned scenarios,\footnote{Alternatively, strong misalignment has also been explored~\cite{Bar-Shalom:2007xeu,Giudice:2011ak}.} where new flavor spurions can be diagonalized in the same flavor basis as the up-quark (or down-quark) Yukawa. In other words, they break the flavor group down to a subgroup larger or equal to the one that leaves the up (or down) Yukawa invariant. In these setups, which have been studied earlier~\cite{Knapen:2015hia,Egana-Ugrinovic:2018znw}, for instance in 2HDM~\cite{Pich:2009sp,Ferreira:2010xe,Botella:2015yfa,Gori:2017qwg,Penuelas:2017ikk,Altmannshofer:2017uvs,Egana-Ugrinovic:2019dqu} or in connection with light scalars and naturalness~\cite{Batell:2017kty,Batell:2021xsi}, chirality-flipping transitions can be generated by new spurions instead of quark masses and therefore might be more likely. A drawback is that, although alignment ensures a CKM suppression of FCNC, it is much weaker than MFV as it gives up the Yukawa suppression. As a result, the NP scale is generically bounded to be well above the TeV scale. Another limitation is the naturalness of the up quark mass, to which the new spurions contribute at loop level. 

In this work, we scrutinize one scenario where the main previous criticism is absent, {\it i.e.}, a flavor-violating scenario in which the strongest constraint on the NP scale comes from measurements at hadron colliders and involves large chirality-flipping processes for the light quark generations. This is achieved upon considering an electroweak (EW) dipole operator in the up-quark
sector of the SMEFT at dimension six, under the assumption of flavor alignment in the same up-quark sector. The dipole operators have a left-right structure, hence they allow one to explore the impact of chirality-flipping NP. Moreover, within the class of left-right operators in the SMEFT at dimension six, dipoles stand out as those which {\it i)} can be probed using low-background events, {\it ii)} can generate energy-growing amplitudes and impact the high-energy tails of kinematic distributions, and {\it iii)} are not suppressed by chirality-flipping spurions in the lepton sector. Furthermore, the focus on the up sector is motivated by the stronger suppression associated to down quark masses. As we demonstrate, our setup generates very mild FCNC, namely the associated bounds are much weaker than those obtained from collider probes. Regarding the latter, we focus on diphotonic $Wh$ productions at hadron colliders, which arise from a dipole operator, have particularly clean backgrounds, and receive negligible contributions from third-generation couplings. Thereby, it is well adapted to the study of large first- or second-generation couplings.

This article is organized as follows. In Sec.~\ref{sec:EFT_framework}, we detail our flavor (and CP) assumption, its realization on the SMEFT, and on the EW dipoles in particular. We also briefly comment on its relation with an alternative flavor assumption, which we dub {\it inverse hierarchy MFV} (IHMFV). In Sec.~\ref{sec:EWdipole_hadron_coll}, we present the projected bounds on the dipole Wilson coefficient from $Wh$ processes at hadron colliders, which we take to be FCC-hh; the results show sensitivity to strongly-coupled UV completions and to the upper edge of weakly-coupled ones. In Sec.~\ref{sec:lebound}, we show that flavor bounds on this setup are much weaker than the collider ones, and we discuss naturalness constraints on quark masses. Finally, we exhibit in Sec.~\ref{sec:simpmodel} a UV-complete model 
that realizes a very particular version of alignment and IHMFV, and 
whose EFT contains EW dipoles as the only flavorful dimension-6 operators. We conclude in Sec.~\ref{sec:conclusion}. 
The work is completed by three appendices. App.~\ref{app:Pheno_analysis} gives details about the collider analysis of Sec.~\ref{sec:EWdipole_hadron_coll}, while App.~\ref{app:pidecay} presents some of the $\chi$PT techniques underlying Sec.~\ref{sec:lebound}. Finally, App.~\ref{app:Oeff} exhibits the whole set of dimension-6 SMEFT operators generated by the model of Sec.~\ref{sec:simpmodel} at one-loop so as to highlight that they do not lead to noticeable flavor-changing effects, and to briefly discuss the additional phenomenology that they drive.

\section{Flavor-Aligned Electroweak Dipoles}
\label{sec:EFT_framework}

In the following, we focus on the collider and flavor effects of the following dimension-6 electroweak quark dipole operators,
\begin{equation}\label{eq:CqW}
    \mathcal{L}_{\rm SMEFT} \supset \frac{c_{uW}}{\Lambda^2} \mcO_{uW}  + \frac{c_{dW}}{\Lambda^2} \mcO_{dW} \text{ with }
    \begin{cases}
    \mcO_{uW} \equiv &W_{\mu\nu}^a\bar{Q}_L\tau^a\sigma^{\mu\nu} u_R\tH\\
    \mcO_{dW} \equiv & W_{\mu\nu}^a\bar{Q}_L\tau^a\sigma^{\mu\nu} d_R H
    \end{cases}  \ ,
\end{equation}
where $\tau^a = \sigma^a/2$, with $\sigma^a$ the Pauli matrices, $\tH=i\sigma^2H^*$, and we omit the flavor indices for conciseness.
More precisely, we assume that only one of these operators is present at a time, mostly the up-quark dipole for definiteness, although very similar collider results apply for the down-quark dipole. We present in Sec.~\ref{sec:simpmodel} an example of a UV model which only generates one of the two dipoles, and also fulfills automatically the flavor assumption that we present now. 

\subsection{Flavor Alignment and CP Quasi-Conservation}
\label{sec:flavor_assumptions}

The Wilson coefficients $c_{uW}$ and $c_{dW}$ are flavorful, {\it i.e.} they have spurious transformations under the global flavor symmetry of the Yukawa-less SM Lagrangian, U$(3)^5=\otimes_{\psi}\text{U}(3)_{\psi}$, that acts on the fermion fields flavor space. Succinctly, each fermion gauge multiplet transforms independently as a fundamental of the associated $SU(3)$ and is the only field in the SM carrying the associated $U(1)$ charge. This flavor symmetry is broken down to its baryon and lepton numbers subgroups $U(1)_B\times U(1)_e\times U(1)_\mu\times U(1)_\tau$ by the SM Yukawas (and to various subgroups by the SMEFT Wilson coefficients), but it can be formally extended to the full SM+dipole Lagrangian, provided the different coupling constants transform as depicted in Tab.~\ref{tab:Yuka_spurions}  (ignoring the abelian factors for conciseness).
\begin{table}[t]
        \centering
        \begin{tabular}{c|c|c|c|c|c}
             & $SU(3)_Q$ & $SU(3)_u$ & $SU(3)_d$ & $SU(3)_L$ & $SU(3)_e$ \\ \hline
            $Y_u$ & $\mathbf{3}$ & $\mathbf{\bar{3}}$ & $\mathbf{1}$ & $\mathbf{1}$ & $\mathbf{1}$ \\[0.1cm]
            $Y_d$ & $\mathbf{3}$ & $\mathbf{1}$ & $\mathbf{\bar{3}}$ & $\mathbf{1}$ & $\mathbf{1}$ \\[0.1cm]
            $Y_e$ & $\mathbf{1}$ & $\mathbf{1}$ & $\mathbf{1}$ & $\mathbf{3}$ & $\mathbf{\bar{3}}$ \\[0.1cm]
            $c_{uW}$ & $\mathbf{3}$ & $\mathbf{\bar{3}}$ & $\mathbf{1}$ & $\mathbf{1}$ & $\mathbf{1}$ \\[0.1cm]
            $c_{dW}$ & $\mathbf{3}$ & $\mathbf{1}$ & $\mathbf{\bar{3}}$ & $\mathbf{1}$ & $\mathbf{1}$
        \end{tabular}
        \caption{Transformation properties of the SM Yukawa couplings and of the electroweak quark dipoles under the (non-abelian) flavor group $SU(3)^5$.}
        \label{tab:Yuka_spurions}
\end{table}
There, we also included the transformations of the SM Yukawas, appearing as follows in the Yukawa sector of the SM,
\begin{equation}
    \mathcal{L}_{\rm Yuk} = - \bar{Q}_L Y_u u_R \widetilde{H} - \bar{Q}_L Y_d d_R H - \bar{L}_L Y_e e_R H + {\rm h.c.} \ .
\end{equation}
Below, restricting to quark quantum numbers, we will also represent the flavor transformations as $Y_u\sim (\mathbf{3}, \mathbf{\bar{3}}, \mathbf{1})$, etc. 

As reminded above, the stringent bounds on flavor-changing processes, particularly on FCNC, restrict the scale of flavor violation or the flavor structure of NP and, in turn, that of the higher-dimensional operators in the SMEFT framework. This applies in particular to the dipoles. Therefore, we make a particular flavor assumption, whose effectiveness will be discussed in Sec.~\ref{sec:lebound}: we consider flavor alignment in either the up or down sector
, depending on which dipole operator we focus on. More precisely, if we focus on the up dipole, we assume that there exists a single new flavor-breaking spurion $\kappa_u$ with the same flavor charges as $Y_u$, and that only $Y_d$ breaks the $U(1)^3$ group that is left invariant by $Y_u$. What that means in practice is that there exists a flavor basis, dubbed the up-basis, in which both $Y_u$ and $\kappa_u$ are diagonal. As we will show below, this assumption is sufficient to suppress FCNC to a satisfactory level.

However, although one could consider extending this assumption to the whole SMEFT, we stress that its effectiveness depends on the dipole being the only operator communicating with the new spurion. For instance, an operator
\begin{equation}
{\cal O}^{(1)}_{qq}=c_{qq,ijkl}^{(1)} \, \bar Q_{L,i} \gamma_\mu Q_{L,j} \, \bar Q_{L,k} \gamma^\mu Q_{L,l}
\end{equation}
with $c_{qq,ijkl}^{(1)}=\left(\kappa_u\kappa_u^\dagger\right)_{ij}\delta_{kl}$, has the appropriate spurion transformation and generates large $\Delta S=1$ FCNC ({\it i.e.}, those which violate strangeness by one unit), unless the eigenvalues of $\kappa_u\kappa_u^\dagger$ along the light generations are much smaller than one. (This is achieved in MFV, where $\kappa_u\propto Y_u$.) A similar statement would apply for $\Delta C=1$ currents (those which violate charmness by one unit), were we considering the down dipole. We will expand more on this in the next section about IHMFV.

Moreover, we need to make an assumption about CP violation (CPV): an arbitrary phase for the diagonal entries of $\kappa_u$ in the up-basis would generate tree-level quark electric dipole moments (EDMs), which are strongly constrained (see, {\it e.g.}, \cite{Pospelov:2005pr,Dekens:2013zca,Panico:2018hal,Kley:2021yhn,Brod:2022bww,Alarcon:2022ero,Falkowski:2023hsg}, and \cite{He:2014xla,Hagiwara:2017ban,Aebischer:2018quc,Aebischer:2018csl,Fuchs:2020uoc,Bahl:2020wee,Bakshi:2021ofj,Degrande:2021zpv,Bonnefoy:2021tbt,Bahl:2022yrs,Kondo:2022wcw,Bonnefoy:2023bzx,Barger:2023wbg} for more discussions of CPV in the SMEFT). This is due to the strong suppression of CPV in the SM \cite{Kobayashi:1973fv,Jarlskog:1985cw,Jarlskog:1985ht}. We therefore assume that, in the absence of $Y_d$, the flavor group is broken to $U(1)^3$ {\it and} CP is preserved. In other words, all CP-odd flavor invariants formed out of $Y_u$ and $\kappa_u$ vanish. Furthermore, we assume that there are no flavor-singlet CP-odd spurions.

Finally, let us stress that our flavor assumption does not imply that $Y_u$ and $c_{uW}$ are diagonal in the same up basis, {\it i.e.} that $c_{uW}=\kappa_u$, but it does imply that they almost are. For instance, assuming a polynomial expansion along the spurions,
\begin{equation}
\label{eq:dipoleAligned}
c_{uW}=P(X_u,X_d)\kappa_u +Q(X_u,X_d)Y_u \ ,
\end{equation}
where $P,Q$ are polynomials of arbitrary degree and $X_u\equiv Y_uY_u^\dagger,\,\,X_d\equiv Y_dY_d^\dagger$, the presence of $Y_d$ slightly misaligns $c_{uW}$ from $Y_u$ and $\kappa_u$. Nevertheless, our claim regarding FCNC holds, as will be explained later. Furthermore, since the off-diagonal entries of $X_d$ in the up basis are suppressed by both CKM elements and down-type Yukawas, they are very subdominant in the collider results that we present later. Therefore, we often neglect them in what follows and take $c_{uW}=\kappa_u$, so that, in the up basis,
\begin{equation}\label{eq:diagYCu}
    Y_u=Y_u^D\;, \quad Y_d=\vckm Y_d^D\;, \quad c_{uW}=c_{uW}^D
    \ ,
\end{equation}
where $\vckm$ is the CKM matrix, $Y_u^D=\text{diag}\left(y_u,y_c, y_t\right)$, $Y_d^D=\text{diag}\left(y_d,y_s, y_b\right)$,   $c_{uW}^D=\text{diag}\left(c_{uW}^{ii}
\right)$. To conclude this section, it is important to note that discussions on alignment can be generalized to 
other
models, {\it e.g.} to the scalar singlet extension of the SMEFT for which flavor alignment was considered in Refs.~\cite{Batell:2021xsi, Batell:2017kty}. We defer the detailed investigation of such extensions for future work.

\subsection{An Alternative: Inverse Hierarchy MFV}\label{sec:IHMFV}

Before presenting the phenomenological analysis of our flavor-aligned assumption on the quark EW dipoles, let us present a closely related assumption, which we refer to as {\it Inverse Hierarchy MFV}. It yields a scenario where NP effects predominantly arise in connection to the first two generations, instead of the third as in MFV, so that the lightest quarks have the largest dipole Wilson coefficients.

MFV~\cite{Chivukula:1987py,Hall:1990ac,Buras:2000dm,DAmbrosio:2002vsn,Isidori:2010kg} 
is defined by requiring that any flavorful coefficient in the SMEFT, {\it i.e.} any Wilson coefficient associated with a higher-dimensional operator that is not a singlet under the flavor symmetry, has to be built out of linear combinations of products of Yukawa matrices to make the overall operator formally (or spuriously) flavor-invariant. For the EW quark dipole operators, it is realized when one sets $P=0$ in Eq.~\eqref{eq:dipoleAligned}. As a consequence, flavor-changing processes are suppressed by CKM entries and SM quark Yukawas. For operators with odd powers of a given fermion multiplet, flavor-changing processes $i\to j$ are suppressed at least by the largest of the Yukawas $y_{ij}$, as well as some more Yukawas (which could be ${\cal O}(1)$ for the top quark) and CKM entries. This is easily seen for the dipole from Eq.~\eqref{eq:dipoleAligned} and \eqref{eq:diagYCu} with $\kappa_u=0$, and it implies that $c_{uW}^{11}\ll c_{uW}^{22}\ll c_{uW}^{33}$ (and similarly for the down case). It is a particular case of the flavor alignment discussed in the previous section.

Another definition of MFV which is sometimes used is that the Yukawa matrices are the only spurions of the flavor symmetry. However, although the above MFV prescription certainly abides by this criterion, it does not exhaust all possibilities. In particular, one is now allowed to consider inverse powers of the Yukawa matrices: defining $\widetilde{Y}\propto\left(Y^\dagger\right)^{-1}$ we find that those spurions follow the same transformation rules as the original Yukawas,
\begin{equation}\label{eq:invY}
 		\widetilde{Y}_u\sim (\mathbf{3}, \mathbf{\bar{3}}, \mathbf{1})\;,\quad \widetilde{Y}_d\sim (\mathbf{3}, \mathbf{1}, \mathbf{\bar{3}})\;,
\end{equation}
so that one can consider flavor-aligned ans\"atze where $\kappa_u=\widetilde{Y}_u$. This observation leads to IHMFV as a generalization of MFV, where one relaxes the (often implicit) assumption that only positive powers of the Yukawas are allowed.

The normalization of $\widetilde{Y}_u$ is chosen in order to have a controlled spurion expansion, {\it i.e.}
$\widetilde{Y}_u = \diag \left(1,\frac{y_u}{y_c},\frac{y_u}{y_t} \right)$ in the up basis.
The IHMFV expansion can also be understood as the MFV expansion where polynomials such as $P,Q$ in Eq.~\eqref{eq:dipoleAligned} are generalized to rational functions. Its UV interpretation is clear: the inverse of flavorful spurions in an EFT can only be obtained upon integrating out heavy particles whose spectrum is dictated by said flavorful spurions. This was noted previously in~\cite{Grinstein:2010ve}, in the context of gauged models of flavor 
where the Yukawa spurions correspond to vacuum expectation values of flavon fields which break the flavor symmetry. Hence, IHMFV captures the low-energy effects of a MFV-respecting theory with heavy fields whose mass spectrum follows a hierarchy similar to that of the SM quarks, dictated by the same UV spurions.\footnote{Due to renormalization group (RG) running, we cannot claim that fields with different interactions --here, the new heavy fields and the SM ones-- maintain a spectrum dictated by the exact same spurions at all scales. If the spurions are generated by the spontaneous breaking of the flavor group, this generically only holds at the breaking scale. However, RG running will not change the hierarchies, which is what we care about here. Therefore, we use a single spurion at all scales in this paper.} This differs from regular MFV, which is restricted to MFV-respecting UV physics which is flavor-blind, or with flavorful interactions and mixings with the SM but with a mass matrix which is a flavor singlet. Then, the normalization of $\tilde Y_u$ which we chose corresponds to normalizing the EFT cutoff $\Lambda$ in Eq.~\eqref{eq:CqW} to the smallest mass scale of this hierarchical spectrum, which is the scale of NP in such models.\footnote{For a spectrum of masses given by the entries of $Y_u M$ for $M$ a UV scale, the lowest NP scale $\Lambda$ is $\sqrt{\text{smallest eigenvalue of }Y_uY_u^\dagger} \times M$, and interactions are suppressed by $(Y_u^\dagger)^{-1}/M=\tilde Y_u/\Lambda$, justifying our normalization of $\tilde Y_u$. $\widetilde{Y}_u=\left(Y_u^\dagger\right)^{-1} \times \sqrt{\text{smallest eigenvalue of }Y_uY_u^\dagger}$ is also the appropriate flavor-covariant definition. One can further make sense of it 
from perturbative unitarity, which tells us that the true cut-off of the EFT should roughly be the scale $\Lambda$ suppressing higher-dimensional operators divided by the largest coupling to the appropriate (rational) power (which is the square root at dimension 6). Therefore, large flavorful couplings such as $(Y_u^\dagger)^{-1}$ would bring the true cutoff much below $\Lambda$. The chosen normalization of $\tilde Y_u$ ensures that the spurion expansion does not diverge and that $\Lambda$ and the true cutoff match.
} The aforementioned UV model of Sec.~\ref{sec:simpmodel} not only fulfills flavor alignment as we defined it, but also exemplifies the UV origin of IHMFV.

For the quark EW dipole operators, IHMFV can generate scalings such as
\begin{equation}\label{eq:cuvihmfv}
	c_{uW}^{D}\propto {\rm diag}(m_u^{-1}, m_c^{-1}, m_t^{-1})
\end{equation}
if $c_{uW}\propto \widetilde{Y}_u$. In this case, the Wilson coefficients matrix is diagonal with the inverse hierarchy of diagonal entries with respect to MFV, $c_{uW}^{11}\gg c_{uW}^{22}\gg c_{uW}^{33}$. IHMFV therefore corresponds to flavor alignment with a specific inverse hierarchy, instead of arbitrary diagonal entries.\footnote{A UV theory following the MFV hypothesis that gives rise to IHMFV couplings in the IR likely also generates EFT coefficients which follow the usual MFV expansion. It is therefore interesting to study the interplay of both contributions with $\mathcal{O}(1)$ flavor-blind coefficients to understand the resulting flavor structure of the EFT. For the dipole, one finds that $c_{uW}^{22} \ll c_{uW}^{11} \text{ or } c_{uW}^{33}$, or both if the scales of flavor-blind and flavorful NP are similar. We will leave this for future study and focus on dipole Wilson coefficients with pure inverse hierarchy.}

Similarly to the case of flavor alignment, when we study the consequences of IHMFV in full generality, we find that it is not very efficient at suppressing FCNC, in the sense of relaxing the scale associated with flavor-violating NP to around the TeV scale, despite standing on similar theoretical grounds as regular MFV. The reason, sketched in~\cite{Grinstein:2010ve}, lies in the absence of Yukawa suppression for the lightest generations. To see this, let us write down a spurionic expansion for the Wilson coefficients in the same way as MFV. We find, for instance, for the spurionic expansion of a chirality-flipping current $c_{Qu,ij} \bar{Q}_i \Gamma u_j$ and a chirality-preserving current $c_{QQ,ij} \bar{Q}_i \Gamma^\prime Q_j$
\begin{equation}
\begin{split}
    c_{Qu} & =  \left( c_1^{Qu} \mathbb{1} + c_2^{Qu} \tilde X_u + c_3^{Qu} \tilde X_d + \dots \right) \tilde Y_u + \text{MFV terms}\;, \\
    c_{QQ} & = \left( c_1^{QQ} \mathbb{1} + c_2^{QQ} \tilde X_u + c_3^{QQ} \tilde X_d + \dots \right) + \text{MFV terms} \;,
\end{split} 
\end{equation}
where $\Gamma,\Gamma^\prime$ collect all possible Lorentz structures, the different constants $c_i^{XY}$ are numbers and $\tilde X_u\equiv \tilde Y_u\tilde Y_u^\dagger,\tilde X_d\equiv \tilde Y_d\tilde Y_d^\dagger$. We are therefore led to introduce two measures for FCNC as follows: for $i\neq j$, in the down basis relevant for, {\it e.g.}, kaon oscillations,
\begin{equation}\label{eq:lambdaMFV}
   (\lamfc)_{ij} =
   X_{u,ij} 
    = \begin{pmatrix} 0 & \lambda^5 & \lambda^3 \\ \lambda^5 & 0 & \lambda^2 \\ \lambda^3 & \lambda^2 & 0 \end{pmatrix}\;,\quad  (\invlamfc)_{ij} =
    \tilde X_{u,ij} 
    \approx 
    \begin{pmatrix} 0 & \lambda & \lambda^3 \\ \lambda & 0 & \lambda^4 \\ \lambda^3 & \lambda^4 & 0 \end{pmatrix} \,,
\end{equation}
where $\lambda\approx 0.2$ is the sine of the Cabibbo angle.
The former measure is what is considered in usual MFV, while the latter corresponds to the IHMFV measure. Comparing $\invlamfc$ to $\lamfc$ in Eq.~\eqref{eq:lambdaMFV}, we see that the size of flavor changing currents is drastically changed. In particular, the `12' elements are less suppressed than in MFV, $\lambda^4 (\invlamfc)_{12} \sim (\lamfc)_{12}$, while the `23' elements are more suppressed in IHMFV, $(\invlamfc)_{23} = \lambda^2 (\lamfc)_{23}$. Therefore, one has to forbid that the IHMFV spurion communicates with most SMEFT operators, in particular those mediating left-handed FCNC, if one wants to evade flavor bounds from meson systems including the first 2 generations of quarks and keep the flavor-violating NP scale at the TeV scale. The model of Sec.~\ref{sec:simpmodel} achieves this, by naturally generating a flavor-violating dipole in the up-sector from up-sector UV physics, together with other flavor-blind SMEFT operators. We will therefore work under the assumption that such a scenario is at play. 

Which dipole one considers in conjunction with a given inverse Yukawa IHMFV spurion also matters.  For instance, using $\tilde Y_d$ together with the down-type electroweak dipole, forming {\it e.g.} $c_{dW}=X_u \tilde Y_d$ would generate tree-level $\Delta S=1$ FCNC such as $K\to\pi\pi\gamma$, only suppressed by $(\vckm{}_{,32})^*\vckm{}_{,31}\sim \lambda^5$. On the other hand, $\Delta C=1$ processes driven by the up dipole with IHMFV in the up sector enjoys the down-type MFV suppression and are suppressed by $(\vckm{}_{,32})^*\vckm{}_{,31}m_b^2\sim \lambda^{11}$. Therefore, up-sector IHMFV, which arises if the MFV-preserving NP is only chiral with respect to $SU(3)_u$ but not $SU(3)_d$, in conjunction with the up dipole is best suited to avoid FCNC. Similar statements hold in the flavor-aligned case, upon comparing the impact of up- or down-sector new spurions. Hence, we focus on the up-quark dipole in what follows.
Furthermore, as for the flavor-aligned case, we need to enforce that CP is not broken by new spurions, which means here that the flavor-blind coefficients of the IHMFV spurion expansion have to be real.

Finally, we stress that one notorious problem of MFV, namely the fact that $y_t \approx 1$ spoils the convergence of the spurion expansion, also applies for IHMFV since $\tilde Y_{u,11}=1$. A treatment along the lines of Refs.~\cite{Feldmann:2008ja,Kagan:2009bn}, or involving an approximate $U(2)$ symmetry~\cite{Agashe:2005hk,Barbieri:2012uh,Isidori:2012ts} (but now acting on the two heaviest quark generations), is possible.\footnote{When both $Y_u$ and $\tilde Y_u$ are present, the large $Y_{u,33}$ and $\tilde Y_{u,11}$ entries only leave a subgroup $U(1)_{Q_2}\times U(1)_c$ unbroken in the up-sector, so that spurions should be assigned transformations under this group instead of a larger one like $U(2)^3$. UV assumptions can nevertheless allow one to use non-abelian approximate symmetries at the matching scale, for instance if the up Yukawa dictates mass hierarchies in a heavy sector which couples to the SM through flavor-blind interactions only. Treating $Y_d$ as a whole as a small spurion, one can expand the flavorful SMEFT Wilson coefficients at the matching scale using three spurions transforming under $U(2)_Q \times U(2)_u \times U(1)_u\times U(3)_d$, where $U(1)_u$ acts identically on $Q_1$ and $u_1$: $\Sigma_d$ in $(\mathbf{2},\mathbf{1},0,\mathbf{\bar 3})$, $\Lambda_d$ in $(\mathbf{1},\mathbf{1},0,\mathbf{3})$ and $\Delta_u$ in $(\mathbf{2},\mathbf{2},0,\mathbf{1})$, which can be chosen so that, in an appropriate flavor basis,
\begin{equation*}
\tilde Y_u=\left(\begin{matrix}1&0\\0&\Delta_u\end{matrix}\right) \ , \quad Y_d=\left(\begin{matrix}\Sigma_d\\\Lambda_d^\dagger\end{matrix}\right) \ .
\end{equation*}
 In particular, in the up basis, $\Delta_u=\text{diag}(m_u/m_c,m_u/m_t)$.}

\section{Electroweak Dipole Operators at Hadron Colliders}
\label{sec:EWdipole_hadron_coll}

Electroweak quark dipole operators are known to generate amplitudes with distinctive kinematical behaviors, facilitating the design of observables sensitive to their presence. In particular, dipole operators tend to generate strong growth with energy. However, this class of operators is often neglected in studies of SMEFT effects at hadron colliders due to their MFV suppression. In Sec.~\ref{sec:IHMFV}, we have argued that UV theories satisfying the MFV assumption may lead to EFTs satisfying IHMFV, such that lighter quark dipole operators are enhanced instead of suppressed. With such UV completions in mind, we chose the flavor assumption presented in Sec.~\ref{sec:flavor_assumptions}, such that chirality-flipping operators are not necessarily suppressed by the appropriate quark mass.  Hence, we revisit the possibility of bounding the Wilson coefficients (WCs) of EW quark dipole operator by looking at the tail of the differential cross-sections at hadron colliders.

\subsection{Growing Dipole Amplitudes}

As a probe of the EW quark dipoles, we focus on diboson processes that are sensitive to the contact interaction induced by these operators. In $Vh$ production in particular, where usually the effect of EW dipoles is neglected by invoking MFV, their contact interactions lead to strong energy growth and hence could be well probed. The same dipole operators also generate growing amplitudes for the Drell-Yan process but with a milder growth since there remains a tree-level propagator and a suppression by the center of mass energy $\sqrt{s}$~\cite{Boughezal:2021tih}, unlike that of 4-fermion operators which are stringently bounded. Generically, we expect those additional operators to be present, since the flavor-aligned ansatz does not restrict the kind of operator that should be included in a generic EFT analysis from the beginning. It would at most justify an analysis with diagonal but
non-universal WCs~\cite{Allwicher:2022gkm}. (As we argued above, all operators need not communicate with the
new aligned spurions.) In Sec.~5, we actually discuss explicit models realizing our flavor-aligned scenario at the level of the dipole operator, and it turns out that they also generate 4-fermion operators at one-loop level. Such additional operators might drive the most stringent bounds on the models that realize our flavor assumption. Nevertheless, we wish to study
the effects of non-universal flavor-diagonal EW quark dipole operators on the phenomenology at hadron colliders, which justifies our focus on V h production as a representative example. We dedicate the rest of this section to this topic.

$Vh$ production is conveniently split into $Wh$ and $Zh$ production. Although both processes share many similarities, the second one is typically affected by more operators and one needs to consider 2 $Z$ decay channels, $Z\to\ell^{+}\ell^{-}$ and $Z\to\nu\bar\nu$, to obtain similar sensitivity as from $Wh$~\cite{Bishara:2020pfx,Bishara:2022vsc}. $Zh$ production at hadron colliders is induced by quarks of the 3 different generations, with the $b\bar b$ contribution being relevant for SMEFT analyses already at LHC~\cite{Rossia:2023hen} and representing $\sim 5\%$ of the cross-section at FCC-hh~\cite{Bishara:2020pfx}. On the other hand, $Wh$ can be produced only by quarks of the first 2 generations. This makes it more suitable to probe the hierarchy between light and heavy quark dipole operators since it can be used to measure the former without contamination of the latter. Heavy quark dipole operators, such as $c_{uW}^{33}$, can be independently constrained in top-quark processes~\cite{Ellis:2020unq,Ethier:2021bye}. 
Thus, we study only the effects of EW dipoles on $Wh$ production.

The operators $\mcO_{uW}$ and $\mcO_{dW}$ generate contact interactions between the quarks, the $W$, and Higgs bosons. This leads to a quadratic growth with energy in the amplitude for hadronic $Wh$ production. More precisely, the part of the dipole operators that generate the contact term follows the general structure,
\begin{equation}\label{eq:dipstruc}
	\mathcal{O}\supset h\partial_\mu W_\nu \bar{q}'\sigma^{\mu\nu}q\;.
\end{equation}
The amplitude of $pp\rightarrow Wh$ can be written in general as,
\begin{equation}
	\mathcal{M}(1_q, 2_{\bar{q}}, 3_W, 4_h)=\sum D\cdot C\cdot \mathcal{T}(h_q, h_{\bar{q}}, h_W)\;,
\end{equation}
where $D$ contains the involved couplings, $C$ encodes the color structure, and $\mathcal{T}$ represents the kinematics dependence of the amplitude. The helicity structure of the dipole operator forces, in the massless quark limit, the polarization of the quark and anti-quark to be the same and this prevents the interference with the SM amplitudes~\cite{Azatov:2016sqh}.
In the all-incoming convention and for massless quarks, the final result is
\begin{equation}
    \mathcal{T}(\pm\frac{1}{2},\pm\frac{1}{2},\pm 1)=\mp \frac{e^{-i\theta}}{\sqrt2} \sin\theta\left(s-m_h^2-m_W^2\right) \left( 1 \pm \sqrt{1-\frac{4\,s\,m_W^2}{(s-m_h^2+m_W^2)^2}} \right),\; \mathcal{T}(\pm\frac{1}{2},\pm\frac{1}{2},0) = 0\;,
\end{equation}
where $\theta $ is the scattering angle and we have assumed that the quarks are approximately massless making the amplitude for the second helicity configuration vanish.

During the rest of this section, we study how $Wh$ production can be used to probe $\mcO_{uW}$ at hadron colliders as the showcase scenario.
The analogous down-type operator, $\mcO_{dW}$, can also be constrained via the same analysis. Since we expect similar results for both operators, we focus on  $\mcO_{uW}$ as a proof-of-concept. Moreover, the up dipole is less constrained from FCNC in the flavor-aligned scenario in the up sector than the down dipole is when flavor alignment occurs in the down sector, as argued in Sec.~\ref{sec:IHMFV}.

\subsection{$Wh$ Production at Hadron Colliders}

In recent years, $Wh$ has been identified as a powerful indirect probe of NP effects at present and future hadron colliders. It shows a high sensitivity to the dimension-6 SMEFT operator $\mcO_{\varphi q}^{(3)}$ thanks to the induced energy growth~\cite{Liu:2018pkg,Baglio:2020oqu,Banerjee:2019twi,Bishara:2022vsc}. Such an effect can be leveraged with a simple binning in the Higgs transverse momentum. Additional angular binning can help also to probe subleading CP-odd operators~\cite{Bishara:2020vix}.

These studies can be carried out already at LHC thanks to the use of the $h\to b\bar b$ decay channel. However, the ideal scenario lies in the future, since FCC-hh would allow to study $Wh$ production at high energies in the $h\to\gamma\gamma$ channel. This final state offers a simpler reconstruction of the Higgs boson and, more importantly, a smaller background than the hadronic decay channels. Hence, it is the ideal place to look for deviations from the SM on high-energy tails.

As a proof-of-concept of how well EW dipole operators can be probed at hadron colliders once the MFV suppression is lifted, we take the analysis of the $Wh\to \ell \nu \gamma\gamma$ process at FCC-hh from Ref.~\cite{Bishara:2020vix}, extend it by computing the dependence of the cross-section on the EW dipole operators and then compute projected bounds on them. During the rest of this section, we describe the main features of the aforementioned analysis, while the details are collected in App.~\ref{app:Pheno_analysis}.

At FCC-hh, the main background processes to $Wh\to \ell \nu \gamma\gamma$ are $W\gamma\gamma$, $W\gamma j$ and $Wjj$ production, with the jet being misidentified as a photon in the latter 2 cases. We assume a conservative jet-to-photon fake rate of $10^{-3}$ and even in that case, the leading background is $W\gamma\gamma$. The main background and the signal were simulated with $0+1$-jet merged samples in order to include the main NLO QCD corrections, which are not negligible. The subleading backgrounds were simulated at LO. We included parton shower and detector simulation effects by using \textsc{Pythia8} and \textsc{Delphes}, the latter with the FCC-hh run card. Generation-level cuts and further details can be found in App.~\ref{app:Pheno_analysis}.

The simple cut-based analysis from Ref.~\cite{Bishara:2020vix} aims at reducing the background cross-section, in particular at high energies. The most effective cuts for this task are the cut on the invariant mass of the photon pair to force it to be around the Higgs mass and a cut on the maximum $p_T$ of the $Wh$ system, which reduces the large contributions from $W\gamma\gamma$ with an additional hard jet~\cite{Bishara:2020vix}. Details about the acceptance and selection cuts of this analysis can be found in App.~\ref{app:Pheno_analysis}. The events were binned according to the $p_T$ of the reconstructed Higgs boson, \pth. Additionally, a second binning on the azimuthal angle of the leptons originating from the $W$ was used. The main goal of this second binning is to allow the measurement to be sensitive to CP-odd operators, but we keep it since it improves the sensitivity to CP-even operators by reducing the impact of systematic uncertainties. The chosen limits of the bins can be found in Tab.~\ref{tab:double_bin}.

\begin{table}[htb!]
    \centering
    \begin{tabular}{|c|c|}
    \hline
     Variable    & Bin limits  \\
     \hline
      $p_T^{h}$   & $\lbrace 200,\, 400,\, 600,\, 800,\, 1000,\, \infty \rbrace$~GeV\\
      $\phi_W$    &  $[-\pi,\,0],\,[0,\,\pi]$\\
         \hline
    \end{tabular}
    \caption{Variables and limits of the bins used in the analysis of the $Wh$ process.}
    \label{tab:double_bin}
\end{table}

\begin{figure}[ht!]
        \centering
        \includegraphics[width=0.74\linewidth]{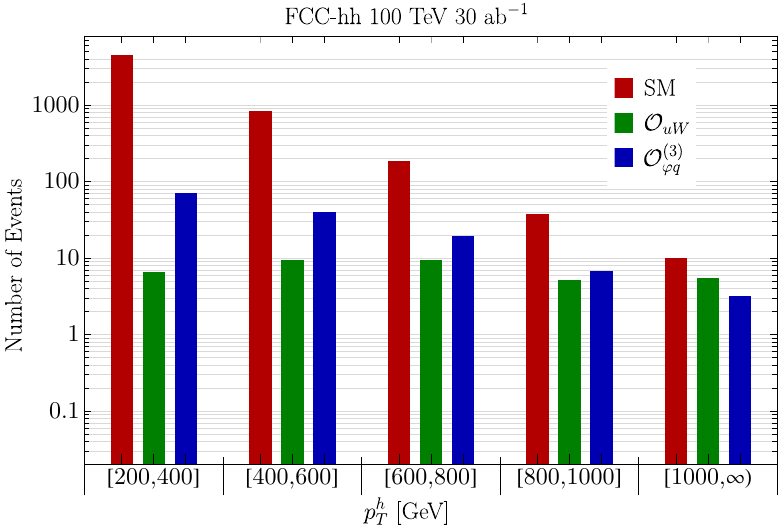}
        \caption{Number of SM and SMEFT events per $p_T^{h}$ bin after selection cuts for the signal and backgrounds at the FCC-hh assuming 30 ab$^{-1}$. The number of SMEFT events is obtained at the upper bound of the corresponding Wilson coefficients from a single operator fit with 5\% syst., {\it i.e.} $c_{\varphi q}^{(3)}= 3\times 10^{-3}$ and $c_{uW}^{11}=1.1\times 10^{-2}$ with $\Lambda = 1$~TeV. Notice that the contribution of ${\cal O}_{\varphi q}^{(3)}$ includes an interference with the SM and could be in principle of either sign. Here we chose the sign such that the interference is positive, {\it i.e.} it adds events to the SM prediction.}
        \label{fig:pThv2}
\end{figure}

We show in Fig.~\ref{fig:pThv2} the number of events at FCC-hh from the $Wh$ process after all the selection cuts in each \pth~bin. We show the contributions of the SM, as well as the ones from the $\mcO_{\varphi q}^{(3)}$ and $\mcO_{u W}$ operators with WCs fixed at the values $c_{\varphi q}^{(3)}= 3\times 10^{-3}$, assuming a flavor-blind operator, and $c_{uW}^{11}=1.1\times 10^{-2}$ with $\Lambda = 1$~TeV, which are representative values of the bounds to be shown in the next section. The contribution from $\mcO_{\varphi q}^{(3)}$ includes its interference with the SM.
The different behavior with energy generated by the $\mcO_{\varphi q}^{(3)}$ and $\mcO_{u W}$ operators can be easily appreciated. This indicates that, when probing the dipole operator, the bound comes from higher-energy bins than in the case of $\mcO_{\varphi q}^{(3)}$.
The total background is smaller than the SM signal in all \pth~bins, as can be seen from Tab.~\ref{tab:sigma_full} in the appendix.

\subsection{Bounds on $\mathcal{O}_{uW}$ from $Wh$ Production}

The analysis outlined in the previous subsection allows us to estimate the FCC-hh sensitivity to dipole operators.
We show in Tab.~\ref{tab:bounds_summary} the projected bounds on the up-quark EW dipole WC, $c_{uW}$, at FCC-hh with $30$~ab$^{-1}$ of integrated luminosity. 
The first two rows show the bound in the diagonal flavor-aligned scenario, in which $c_{uW}^{11}$ and $c_{uW}^{22}$ are independent WCs.  
In the third row, we show the result for a light-flavor-universal scenario in which $c_{uW}^{11}=c_{uW}^{22}=c_{uW}$. We do not specify the top-quark WCs, which are associated with negligible effects due to the very small top content of the proton. In our scenario, it would be best constrained by flavor data (see footnote~\ref{footnote:topDipoleFlavor}). The last row shows the bounds from this analysis on the flavor-universal $c_{\varphi q}^{(3)}$, the WC that this process is most sensitive to. The right column shows the bound from a one-operator fit, while the middle column shows the result of profiling over the other WCs in the fit. For the first two rows, this means profiling over $c_{\varphi q}^{(3)}$ and the other dipole WC, while for the flavor-universal case, it is profiling over $c_{\varphi q}^{(3)}$. Notice that in the IHMFV scenario, the dipole coefficients are related as $c_{uW}^{11}/c_{uW}^{22}=m_c/m_u\approx 797$, and hence the corresponding bounds would be approximately equal to the ones on $c_{uW}^{11}$. 

The projected bounds on $c_{uW}^{11}$ are almost a factor of $4$ worse than the bounds on $c_{\varphi q}^{(3)}$. This was expected from the lack of interference between the dipole operator amplitude and the SM one, and since only one quark flavor, instead of four, contributes to the $c_{uW}^{11}$ bound. The sensitivity to $c_{uW}^{22}$ is a further factor of $\sim 3$ worse due to the lower content of second-generation quarks of the proton. Hence, the flavor-universal results are almost identical to the ones for $c_{uW}^{11}$. The hierarchy in the bounds between $c_{\varphi q}^{(3)}$ and $c_{uW}$ ensures that profiling has a limited impact on the $c_{\varphi q}^{(3)}$ bounds. We checked that using the second binning in $\phi_W$ also reduces such impact. For a detailed analysis of the sensitivity to $c_{\varphi q}^{(3)}$, see Ref.~\cite{Bishara:2020vix}.

\begin{table}
\begin{centering}
\begin{tabular}{c|c|c}
\toprule
Coefficient & Profiled Fit & One Operator Fit \tabularnewline
\hline 
$c_{uW}^{11}$ &
\begin{tabular}{ll}
\rule{0pt}{1.25em}$[-1.33,\,1.33]\times10^{-2}$ & $1\%$ syst.\\
\rule{0pt}{1.25em}$[-1.39,\,1.39]\times10^{-2}$ & $5\%$ syst.\\
\rule[-.65em]{0pt}{1.9em}$[-1.47,\,1.47]\times10^{-2}$ & $10\%$ syst.
\end{tabular}
&
\begin{tabular}{ll}
\rule{0pt}{1.25em}$[-1.11,\,1.11]\times10^{-2}$ & $1\%$ syst.\\
\rule{0pt}{1.25em}$[-1.12,\,1.12]\times10^{-2}$ & $5\%$ syst.\\
\rule[-.65em]{0pt}{1.9em}$[-1.15,\,1.15]\times10^{-2}$ & $10\%$ syst.
\end{tabular}
\tabularnewline

\hline
$c_{uW}^{22}$ &
\begin{tabular}{ll}
\rule{0pt}{1.25em}$[-4.2,\,4.2]\times10^{-2}$ & $1\%$ syst.\\
\rule{0pt}{1.25em}$[-4.5,\,4.5]\times10^{-2}$ & $5\%$ syst.\\
\rule[-.65em]{0pt}{1.9em}$[-4.7,\,4.7]\times10^{-2}$ & $10\%$ syst.
\end{tabular}
&
\begin{tabular}{ll}
\rule{0pt}{1.25em}$[-3.2,\,3.2]\times10^{-2}$ & $1\%$ syst.\\
\rule{0pt}{1.25em}$[-3.3,\,3.3]\times10^{-2}$ & $5\%$ syst.\\
\rule[-.65em]{0pt}{1.9em}$[-3.4,\,3.4]\times10^{-2}$ & $10\%$ syst.\\
\end{tabular}
\tabularnewline

\hline
$c_{uW}$ &
\begin{tabular}{ll}
\rule{0pt}{1.25em}$[-1.27,\,1.27]\times10^{-2}$ & $1\%$ syst.\\
\rule{0pt}{1.25em}$[-1.34,\,1.34]\times10^{-2}$ & $5\%$ syst.\\
\rule[-.65em]{0pt}{1.9em}$[-1.40,\,1.40]\times10^{-2}$ & $10\%$ syst.
\end{tabular}
&
\begin{tabular}{ll}
\rule{0pt}{1.25em}$[-1.05,\,1.05]\times10^{-2}$ & $1\%$ syst.\\
\rule{0pt}{1.25em}$[-1.07,\,1.07]\times10^{-2}$ & $5\%$ syst.\\
\rule[-.65em]{0pt}{1.9em}$[-1.09,\,1.09]\times10^{-2}$ & $10\%$ syst.\\
\end{tabular}
\tabularnewline

\hline
$c_{\varphi q}^{(3)}$ &
\begin{tabular}{ll}
\rule{0pt}{1.25em}$[-4.6,\,2.5]\times10^{-3}$ & $1\%$ syst.\\
\rule{0pt}{1.25em}$[-6.3,\,3.0]\times10^{-3}$ & $5\%$ syst.\\
\rule[-.65em]{0pt}{1.9em}$[-8.3,\,3.5]\times10^{-3}$ & $10\%$ syst.
\end{tabular}
&
\begin{tabular}{ll}
\rule{0pt}{1.25em}$[-2.7,\,2.5]\times10^{-3}$ & $1\%$ syst.\\
\rule{0pt}{1.25em}$[-3.3,\,2.9]\times10^{-3}$ & $5\%$ syst.\\
\rule[-.65em]{0pt}{1.9em}$[-4.0,\,3.5]\times10^{-3}$ & $10\%$ syst.
\end{tabular} \\
\bottomrule

\end{tabular}
\par\end{centering}
\caption[caption]{Bounds at $95\%$ C.L.~on the coefficients of the $c_{uW}^{11}$, $c_{uW}^{22}$, $c_{uW}$(flavor-universal case with $c_{uW}^{11}=c_{uW}^{22}$), and $c_{\varphi,q}^{(3)}$ setting $\Lambda = 1 \, \text{TeV}$. {\bf Left column:} bounds profiling over the other coefficients. {\bf Right column:} bounds with a one operator fit, {\it i.e.} setting the other two coefficients to zero. For the flavor-universal case, the bounds of $c_{uW}$ is obtained by profiling over $c_{\varphi q}^{(3)}$ or set it to zero. The profiled bound on $c_{\varphi q}^{(3)}$ was obtained in the case of free $c_{uW}^{11}$ and $c_{uW}^{22}$.}
\label{tab:bounds_summary}
\end{table}

\begin{figure}[t]
        \centering
        \includegraphics[width=0.45\linewidth]{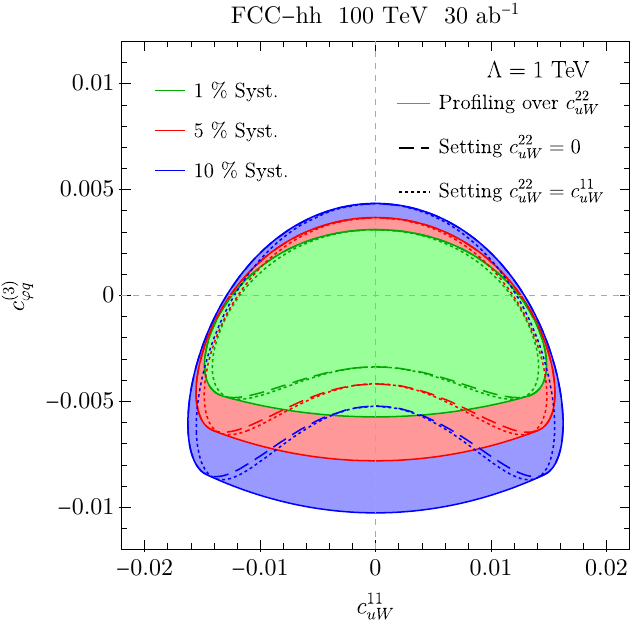} \includegraphics[width=0.45\linewidth]{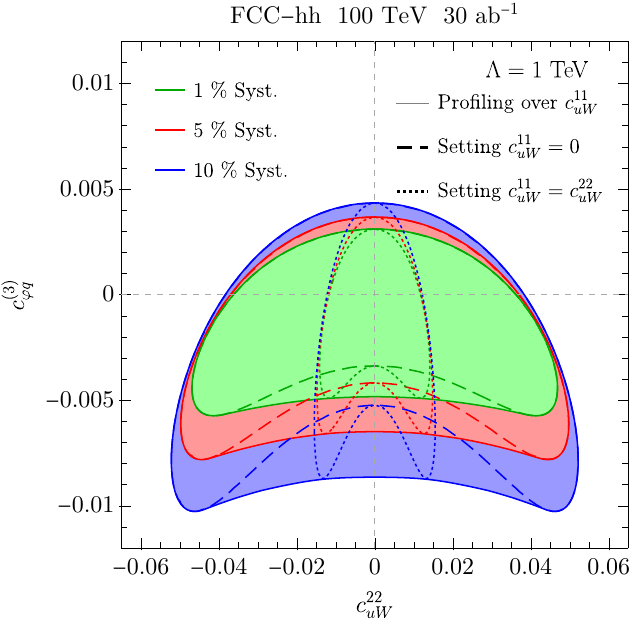}
        \caption{Expected $95\%$ C.L. bounds on $c_{uW}^{11}$, $c_{uW}^{22}$, $c_{\varphi q}^{(3)}$ at the FCC-hh for $30 \, \text{ab}^{-1}$.}
        \label{fig:bounds2d_v2}
\end{figure}

In Fig.~\ref{fig:bounds2d_v2}, we show the $95\%$ C.L. on the planes $c_{\varphi q}^{(3)}-c_{uW}^{11}$ (left panel) and $c_{\varphi q}^{(3)}-c_{uW}^{22}$ (right panel). In both cases, we show the result of profiling over the other dipole WC (full line), setting it to zero (dashed line) or linking it via flavor universality (dotted line). In the latter case, the x-axis should be read as $c_{uW}$. The left panel shows that setting $c_{uW}^{22}=0$ or $c_{uW}^{22}=c_{uW}^{11}$ has a negligible impact since our analysis is mostly sensitive to the first-generation quarks. However, profiling over $c_{uW}^{22}$ explores larger values of $c_{uW}^{22}$ and causes a sizeable difference in the correlation between $c_{\varphi q}^{(3)}$ and $c_{uW}^{11}$ for negative values of the former. The opposite situation can be observed in the right panel, where each choice for $c_{uW}^{11}$ causes very different results, especially for negative $c_{\varphi q}^{(3)}$. In particular, setting $c_{uW}^{11}=c_{uW}^{22}$ generates much tighter bounds.

Had we considered the $h\to b\bar b$ decay channel and adopted the analysis from Ref.~\cite{Bishara:2022vsc}, the projected bounds at FCC-hh would have been of the same order of magnitude, in particular for $5\%$~syst. uncertainty. The projections in Ref.~\cite{Bishara:2022vsc} degrade by a factor of $\sim 5-6$ when going from FCC-hh to HL-LHC.\footnote{We chose the diphotonic $Wh$ analysis at FCC-hh as the showcase due to its simplicity and constraining power.} 
Hence, we estimate that their analysis for the 1-lepton channel, {\it i.e.} $Wh$, could yield the HL-LHC bounds $|c_{uW}^{11}|\lesssim 6\times 10^{-2}$, $|c_{uW}^{22}|\lesssim 2\times 10^{-1}$, and $|c_{uW}|\lesssim 6\times 10^{-2}$ with $\Lambda=1$~TeV. A reduction of the luminosity to the values of Run 2 ($\mathcal{L}=139$ fb$^{-1}$) or 3 ($\mathcal{L}= 300$ fb$^{-1}$) worsens the bounds by $\sim 80\%$ or $\sim 50\%$ respectively. Thus, our estimated bounds for LHC Run 2 (3) are $|c_{uW}^{11}|\lesssim 1.1 (0.9)\times10^{-1}$, $|c_{uW}^{22}|\lesssim 4(3)\times10^{-1}$, and $|c_{uW}|\lesssim 1.1(0.9)\times10^{-1}$ for $\Lambda=1$ TeV.
This should be compared against current and projected bounds from other collider processes. For instance, the quark dipole operator has also been studied with Drell-Yan data from LHC~\cite{daSilvaAlmeida:2019cbr, Boughezal:2021tih,Allwicher:2022gkm}. 
For $\Lambda=1$~TeV, Ref.~\cite{daSilvaAlmeida:2019cbr} obtain $|c_{uW}|<3.8\times 10^{-1}$ from a single-operator fit and $|c_{uW}|<5.3\times10^{-1}$ from a profiled fit, in agreement with Ref.~\cite{Boughezal:2021tih}. Assuming that these bounds are statistically limited, we can rescale them to HL-LHC luminosity and obtain $|c_{uW}|\lesssim 1\times 10^{-1}$ and  $|c_{uW}|\lesssim 2\times 10^{-1}$.
Ref.~\cite{Allwicher:2022gkm} finds slightly better bounds from LHC data, $|c_{uW}^{11}|\lesssim 1.4\times 10^{-1}$ and $|c_{uW}^{22}|\lesssim 5.7\times 10^{-1}$, which could become equal to our $Wh$ estimates at HL-LHC.
A recent study shows that at HL-LHC, $Zh\to\ell^{+}\ell^{-}b\bar{b}$ will have worse sensitivity to $c_{uW}$ than our expectation from $Wh$~\cite{Bhattacharya:2024sxl}.
Overall, at LHC, $Wh$ and Drell-Yan show a similar sensitivity to these operators and further studies are needed for a detailed comparison.
There are no Drell-Yan estimates for FCC-hh that could be easily compared against our bounds.
In addition to high-energy tails, electroweak precision data also constrains the light quark dipole coupling to electroweak gauge bosons; for instance the coefficient $c_{uW}$ will modify the $Z$ decay width. However, these bounds are weak, $|c_{uW}|\lesssim 10$, in part due to the mass-suppressed interference between the SM and dipole amplitudes~\cite{Escribano:1993xr,Kopp:1994qv,daSilvaAlmeida:2019cbr,Brivio:2019myy}.
Therefore, the high-energy regime of $Wh$ could be a useful probe of light-quark EW dipole operators at LHC,
HL-LHC and FCC-hh. 

Finally, it is instructive to assess the EFT validity in this analysis. In agreement with what was found in Ref.~\cite{Bishara:2020vix}, the bounds presented before are valid for an EFT cutoff\footnote{Notice that the scale $\Lambda$ which we use throughout this paper and often fix to $1$~TeV does not capture exactly the UV cutoff of the theory. From the bottom-up, the latter can be bounded via, {\it e.g.}, perturbative unitarity considerations, in which case the upper bound contains $\Lambda$ but also WCs and numerical factors~\cite{Corbett:2014ora,Corbett:2017qgl,Cohen:2021gdw}. From the top-down, the mass scale of NP relates to $\Lambda$ in a way that depends on the UV couplings, see e.g.~\cite{DiLuzio:2016sur}.} $\gtrsim 5$~TeV, while they degrade slightly for a cutoff $\sim2-5$~TeV and the analysis becomes invalid for lower cutoff scales. To properly interpret our bounds in terms of UV physics, it is also worth mentioning that the dipole operators are only generated at the 1-loop level in weakly coupled theories~\cite{Craig:2019wmo}, so that one might expect a naive bound of $c_{uW} \sim \frac{\mathcal{O}(1)}{16\pi^2} \approx 6.3 \cdot 10^{-3}$ which is beyond the reach of most of our bounds. However, we expect a scaling $c_{uW}\sim \frac{g\,g_{*}^{3}}{16\pi^2 M^2}$, where $g_{*}$ is the coupling between UV and SM fields, for a typical UV model. Our bounds would then mean that we are probing $g_{*}\gtrsim 3.9$. Such couplings, despite being large, remain well within the perturbative regime $g_{*}<\,4\pi$~\cite{Giudice:2007fh,Manohar:1983md,Georgi:1986kr}. 

Another marker of a possible EFT validity loss would be a relevant contribution from operators of dimension 8 or higher. Indeed, due to the lack of interference between the SM and dipole amplitudes, dimension-8 operators contribute at the same order, $\Lambda^{-4}$, in the SMEFT expansion. Motivated by the fact that recent studies in the geoSMEFT framework indicate that contributions to $Wh$ production from interference between dimension-8 and SM amplitudes are highly suppressed by the SM couplings~\cite{Corbett:2023yhk}, and by the fact that we would also need to make flavor and operator assumptions at dimension-8, themselves backed by UV model examples as in Sec.~\ref{sec:simpmodel}, we neglect dimension-8 contributions. Nevertheless, a more precise study of their impact on dipole bounds using any of the SMEFT dimension-8 bases would be interesting and is left for future work.

\section{Low Energy Constraints}
\label{sec:lebound}

In this section, we explore the constraints that are imposed by low-energy data on EW quark dipoles in flavor-aligned scenarios. Usually, bounds on contributions to flavor-changing currents beyond the SM involving the first two generations give stringent constraints on off-diagonal flavorful couplings, pushing the NP scale to high values. However, we will show that those constraints loosen significantly in the case of the EW up-quark dipole $\mathcal{O}_{uW}$ in the flavor-aligned scenario when new spurions only appear in the up sector. We will also analyze the particular case of IHMFV. Overall, the flavor bounds turn out to be much weaker than the collider ones.

We remind the reader that, as explained above, we needed to assume the absence of any new CP-odd spurion beyond that of the SM, due to the strong EDM constraints on the imaginary part of dipole operators. Also, if additional SMEFT operators are present, we need to assume that those most restricted by low-energy flavor data, such as purely left-handed 4-Fermi operators, do not communicate with the new flavor-breaking spurions. We also remind that we illustrate how the above assumptions are realized in a UV model in Sec.~\ref{sec:simpmodel}.

In the SM and beyond, translating low-energy flavor bounds to the level of UV models first requires that one integrates weak-scale dynamics out, matching it to the Weak Effective Theory (WET) Hamiltonian~\cite{Jenkins:2017jig,Jenkins:2017dyc,Dekens:2019ept},
\begin{equation}\label{WETlag}
    \mathcal{H}_{\rm eff}=G_F\sum_i C_i\mathcal{Q}_i+{\rm h.c.}\;,
\end{equation}
where $G_F$ is the Fermi constant and $C_i$ are Wilson coefficients. The latter run to lower energies until the QCD scale, where a theory of hadrons, most particularly Chiral Perturbation Theory ($\chi$PT) for the light mesons, takes over. Through this matching and running procedure~\cite{Buchalla:1995vs,Aebischer:2015fzz,Celis:2017hod,Jenkins:2017jig,Jenkins:2017dyc,Aebischer:2018bkb,Hurth:2019ula,Dekens:2019ept,Liao:2020zyx}, one can obtain constraints on the WET WCs and on NP from those on FCNC, arising from various meson experiments. In particular, rare kaon decays are sensitive low-energy probes of BSM physics~\cite{Cirigliano:2011ny}. 

Under our flavor assumption that only the dipole has a non-MFV flavor structure in the SMEFT, tree-level processes are either MFV-suppressed or flavor-changing charged currents (FCCC), 
while FCNC will arise at the loop level. We discuss both types of processes in the following sections. We finally explore the constraints imposed by the naturalness of the quark masses at the weak scale.

\subsection{FCNC }

\begin{figure*}[ht!]
\centering
\begin{tikzpicture}[node distance=1.5cm and 1.5cm]
   \coordinate[label=left:$\bar{s}$] (sb);
   \coordinate[blob, right=of sb] (v1);
   \coordinate[vertex, right=of v1] (v3);
   \coordinate[label=right:$\bar{\ell} (\bar{q})$, right=of v3] (lb);
   \coordinate[label=left:$d$, above=of sb] (d);
   \coordinate[blob, right=of d] (v2);
   \coordinate[vertex, right=of v2] (v4);
   \coordinate[label=right:$\ell(q)$, right=of v4] (l);
   \draw[fermion] (v1)--(sb);
   \draw[fermion] (v2)--(v1);
   \draw[fermion] (d)--(v2);
   \draw[fermion] (lb)--(v3);
   \draw[fermion] (v3)--(v4);
   \draw[fermion] (v4)--(l);
   \draw[photon] (v2)--(v4);
   \draw[photon] (v1)--(v3);
  \end{tikzpicture} 
  \begin{tikzpicture}[node distance=1.2cm and 1.2cm]
   \coordinate[label=left:$\bar{s}$] (sb);
   \coordinate[blob, right=of sb] (v1);
   \coordinate[blob, above=of v1] (v2);
   \coordinate[label=left:$d$, left=of v2] (d);
   \node[] at ( $ (v1)!0.5!(v2) $ ) (vx) {};
   \coordinate[vertex, right=of vx] (v3);
   \coordinate[vertex, right=of v3] (v4);
   \node[label=right:$\ell(q)$] at ( $ (d)!4.0!(v2) $ ) (l) {};
   \node[label=right:$\bar{\ell} (\bar{q})$] at ( $ (sb)!4.0!(v1) $ ) (lb) {};
   \draw[fermion] (v1)--(sb);
   \draw[fermion] (v2)--(v1);
   \draw[fermion] (d)--(v2);
   \draw[photon] (v1)--(v3);
   \draw[photon] (v2)--(v3);
   \draw[photon] (v3)--(v4);
   \draw[fermion] (v4)--(l);
   \draw[fermion] (lb)--(v4);
  \end{tikzpicture} 

\begin{tikzpicture}[node distance=0.8cm and 0.8cm]
   \coordinate[label=left:$\bar{s}$] (sb);
   \coordinate[blob, right=of sb] (v1);
   \coordinate[vertex, right=of v1] (v2);
   \coordinate[blob, right=of v2] (v3);
   \coordinate[label=right:$\gamma (g)$, below=of v3] (v4);
   \coordinate[label=right:$d$, right=of v3] (d);
   \draw[fermion] (v1)--(sb);
   \draw[fermion] (v1)--(v2);
   \draw[photon] (v3) arc[start angle=0, delta angle=180, radius=1.0cm];
   \draw[fermion] (v2)--(v3);
   \draw[fermion] (v3)--(d);
   \draw[photon] (v4)--(v2);
  \end{tikzpicture}
 \begin{tikzpicture}[node distance=0.8cm and 0.8cm]
   \coordinate[label=left:$\bar{s}$] (sb);
   \coordinate[blob, right=of sb] (v1);
   \coordinate[blob, right=of v1] (v2);
   \coordinate[vertex, right=of v2] (v3);
   \coordinate[label=right:$\gamma $, below=of v3] (v4);
   \coordinate[label=right:$d$, right=of v3] (d);
   \draw[fermion] (v1)--(sb);
   \draw[fermion] (v1)--(v2);
   \draw[photon] (v3) arc[start angle=0, delta angle=180, radius=1.0cm];
   \draw[fermion] (v2)--(v3);
   \draw[fermion] (v3)--(d);
   \draw[photon] (v4)--(v2);
  \end{tikzpicture}
\begin{tikzpicture}[node distance=0.8cm and 0.8cm]
   \coordinate[label=left:$\bar{s}$] (sb);
   \coordinate[blob, right=of sb] (v1);
   \coordinate[right=of v1] (v2);
   \coordinate[blob, right=of v2] (v3);
   \coordinate[label=right:$d$, right=of v3] (d);
   \draw[fermion] (v1)--(sb);
   \draw[photon] (v3) arc[start angle=0, delta angle=180, radius=1.0cm]  node[pos=0.25,vertex] (v4) {};
   \node[] at ( $ (v2)!2.5!(v4) $ ) (v5) {$\gamma$};
   \draw[fermion] (v1)--(v3);
   \draw[fermion] (v3)--(d);
   \draw[photon] (v4)--(v5);
  \end{tikzpicture}
\caption{Feynman diagrams with two up-dipole insertions (represented by the large black dots) inducing $\Delta S=1$ WET operators. In the first row, fermion lines on the right side of the diagrams can be either leptons $\ell$ or quarks $q$. In the second row, the external gauge boson in the first diagram could also be a gluon.}
\label{fig:fcncgim}
\end{figure*}
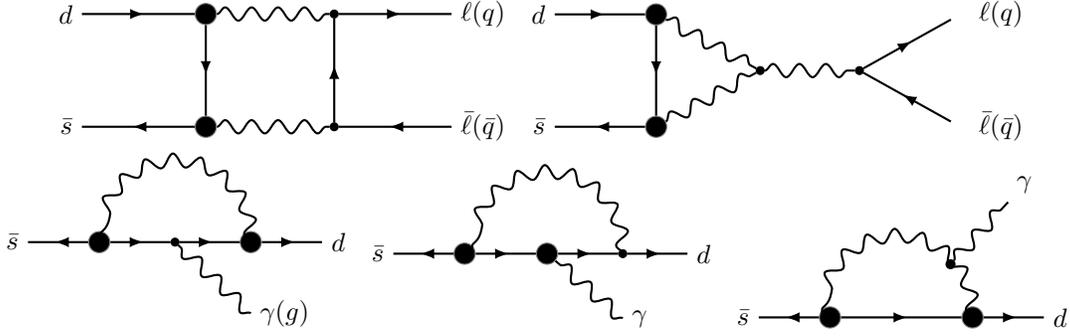

The most stringent flavor bounds usually arise from FCNC, hence we start our analysis from those. Even when they are absent at tree level, some of the FCNC operators in \eqref{WETlag} are generated at loop level by the dipole, as shown in Fig.~\ref{fig:fcncgim}. This figure shows diagrams with $\Delta F=1$ (where $\Delta F=\Delta S,\,\Delta C$) and two insertions of the dipole operator that generate FCNC in the down-quark sector.\footnote{We ignore dimension-8 SMEFT operators, although they may contribute to the processes of interest at the same order in the SMEFT expansion. How our flavor assumption should extend to these operators is left for future work.} 
The flavor suppression with a single insertion of the aligned dipole operator to the considered observables is $y_{u,i} C_{uW,ii} V_{{\rm CKM},ij} V_{{\rm CKM},ik}^{*}$, where $i$ labels the up-type quark running in the loop and $j,k$ label the external quarks. For an inversely hierarchical scenario, $c_{uW} Y_u \propto \mathrm{1}$, therefore 
flavor violation in such diagrams is very suppressed due to 
the GIM mechanism.
In a generic aligned scenario, the contributions from the second and third generations are comparable, and they are a factor of $10^3$ larger than the contribution from the first generation.\footnote{\label{footnote:topDipoleFlavor}Note that a sizeable $c_{uW}^{33}$ is compatible with the regular MFV assumption, and therefore it cannot generate larger flavor-violating effects than the complete SMEFT constrained by that assumption. In that case, low-energy flavor bounds imply that the cutoff should be larger than $\sim$ TeV, even accounting for all possible tree-level operators~\cite{DAmbrosio:2002vsn}. In our case, flavor-violating observables all originate at loop level from the dipole, and are expected to be much weaker when they are driven by $c_{uW}^{33}$. Given the strength of the bounds that we derive below, we conclude that the presence of $c_{uW}^{33}$ does not affect noticeably the low-energy flavor phenomenology. This also allows us to focus on the light-generation dipoles $c_{uW}^{11,22}$ in the current section, so that the results can directly be compared to the collider bounds of Sec.~3, which are only stringent at the level of the two light generations. At colliders, an aligned up-type EW quark dipole with a sizable third generation entry would, for instance, receive stringent bounds from Higgs data (e.g. $H\to\gamma\gamma$). Recent global fits of the LHC Run-2 data suggest bounds similar to or slightly stronger than the ones presented here~\cite{Celada:2024mcf}.} We have checked that they are smaller in magnitude for cutoffs close to the collider bounds. For IHMFV, they are either down-Yukawa-suppressed or flavor-diagonal. Diagrams in the up sector are flavor-diagonal or strongly suppressed by the down Yukawa. Finally, $\Delta F=2$ operators are suppressed by the GIM mechanism and by MFV in the down sector.
Therefore, in this subsection, we focus on WET operators generated by diagrams featuring two dipole insertions and giving rise to $\Delta S=1$ processes (Those are the most sensitive $\Delta F=1$ processes given our flavor assumption.). They are the following,
\bea
\label{eq:WETops}
&\mathcal{Q}_{\ell 1}=(\bar{d}\gamma^\mu P_Ld)(\bar{\ell}\gamma_\mu P_L\ell)\;,\quad \mathcal{Q}_{7\gamma}^{(\prime)}=e m_{d_R} \left(\bar{d}
\sigma^{\mu\nu}P_{R(L)}d\right) F_{\mu\nu} \;,\\
 &\mathcal{Q}_{q3}=\left(\bar{d}\gamma^\mu P_Ld\right)\left(\bar{q}\gamma_\mu P_L q\right)\;,\quad \mathcal{Q}_{8g}^{(\prime)}=g_s m_{d_R} \left(\bar{d}
\sigma^{\mu\nu}P_{R(L)}T^A d\right) G^A_{\mu\nu} \;,
\eea
where $q$ represents any type of quark, $\ell$ is any kind of lepton, and the generation indices are kept implicit. We explicitly pulled a quark mass out of the Wilson coefficients of the magnetic and chromo-magnetic dipoles, since it arises from the diagram in Fig.~\ref{fig:fcncgim}, due to our assumption on the flavor and operator structure in the SMEFT.

They contribute to several flavor-changing processes involving mesons, and we consider here kaon decays to lepton pairs or photon pairs, as well as $K^+\rightarrow \pi^+\pi^0\gamma$. $K\rightarrow\pi\pi$ can also be induced by $\mathcal{Q}_{q3}$ and $\mathcal{Q}_{8g}^{(\prime)}$~\cite{Pich:2021yll}, however, current uncertainties in SM computations via lattice simulations limit the accuracy in determining chiral Lagrangian parameters. Therefore, we leave a discussion of $K\rightarrow\pi\pi$ to App.~\ref{app:kpipi}. For a given process, several operators in Eq.~\eqref{eq:WETops} are likely to contribute at a given order in $\chi$PT. For instance, both the photon and gluon dipoles can contribute to $K \rightarrow \gamma\gamma$ at ${\cal O}(p^6)$, or the 4-quark operator $\mathcal{Q}_{q3}$ can contribute to $K^+\rightarrow\pi^+\pi^0\gamma$ through the inner bremsstrahlung~\cite{good1959pion}. We leave a precise determination of each operator contribution for future work, and focus on that of $\mathcal{Q}_{\ell 1}$ and $\mathcal{Q}_{7\gamma}^{(\prime)}$. We checked that all other contributions to the processes of interest are smaller or equal according to the $\chi$PT power counting, so $\mathcal{Q}_{\ell 1}$ and $\mathcal{Q}_{7\gamma}^{(\prime)}$ serve as a proxy for the full result. As we will show, the resulting bounds are weaker than collider ones, which justifies a posteriori that an order-of-magnitude analysis is sufficient.

The Wilson coefficients of $\mathcal{Q}_{\ell1}$ and $\mathcal{Q}_{7\gamma}^{(\prime)}$ are the following, 
\begin{equation}\label{eq:C4F}
	C_{\ell1}= \frac{z_{\ell 1}c_{uW}c_{uW}^\dagger m_W^2}{16\pi^2 G_F\Lambda^4}\;,\quad C_{7\gamma}=C_{7\gamma}'= \frac{z_{7\gamma} c_{uW}c_{uW}^\dagger v_h^2}{16\pi^2 G_F\Lambda^4}\;,
\end{equation}
where the values of $z_i$ are log-dependent of the renormalization scale $\mu\sim m_K$. We find that the leading contribution of $z_i$ are about $\mathcal{O}(10)$, 
$z_{\ell 1}\approx 3\log (m_W/m_K)+3/2$, and $z_{7\gamma}\approx (10/9)\log(m_W/m_K)-277/54$. Using our flavor assumption from Eq.~\eqref{eq:diagYCu}, if $c_{uW}^{11}$ is not accidentally small, the off-diagonal components of the Wilson coefficients depend on $ \left(c_{uW}c_{uW}^\dagger\right)_{i\neq j} \approx \left|c_{uW}^{11}\right|^2\invlamfc$, which is real at the leading order, so that we can focus on FCNC constraints applicable to the real components of $C_i$. (We nonetheless remind that we have assumed no new source of CP violation.) 

The experimental bounds on $C_{\ell 1}$ can be obtained from kaon leptonic decays. They also receive a contribution from $C_{7\gamma}^{(\prime)}$, but it is suppressed by $\sin^2\theta_{\rm w}m_s/m_K$. 
The most stringent limits on BSM physics arise from $K_L\rightarrow \ell^+\ell^-$, for which the SM prediction has been recently improved~\cite{Hoferichter:2023wiy}.
Normalizing to the $K_L\rightarrow\gamma\gamma$ decay, the decay rate can be expressed in terms of the reduced amplitude $\mathcal{A}_\ell$~\cite{Cirigliano:2011ny},
\begin{equation}
    R_L^\ell =\frac{\BR(K_L\rightarrow\ell^+\ell^-)}{\BR(K_L\rightarrow\gamma\gamma)}=2\sqrt{1-\frac{4m_\ell^2}{m_K^2}}\left(\frac{\alpha_{\rm em}m_\ell}{\pi m_K}\right)^2\left|\mathcal{A}_\ell\right|^2\;.
\end{equation}
Using $\BR(K_L\rightarrow\gamma\gamma)=5.47(4)\times 10^{-4}$~\cite{Workman:2022ynf}, 
the experimental value of $R_L^{\mu}$ implies that $ {\rm Re}\,\mathcal{A}_\mu^{\rm exp}=\pm 1.16(24) $, while the SM prediction is ${\rm Re}\,\mathcal{A}_\mu^{\rm SM}=-1.96(36)$~\cite{Hoferichter:2023wiy}. Assuming the negative value for ${\rm Re}\,\mathcal{A}_\mu^{\rm exp}$, we can derive the BSM constraint ${\rm Re}\,\mathcal{A}_\mu^{\rm BSM}=0.80(43)$.\footnote{We refrain from deriving an analogous constraint from $R_L^{e}$ due to high experimental uncertainties.}
The presence of $\mathcal{Q}_{\ell 1}$ leads to
\begin{equation}
    {\rm Re}\,\mathcal{A}_\ell^{\rm BSM}=\frac{\pi G_F m_K f_K C_{\ell 1}^{ds}}{\alpha_{\rm em}}\sqrt{\frac{m_K}{16\pi\Gamma (K_L\rightarrow\gamma\gamma)}}\;,
\end{equation}
where $f_K=35\,\mev/V_{us}$~\cite{Workman:2022ynf} refers to the meson decay constant. Demanding that the contribution of the WET operator ${\cal Q}_{\ell 1}$ does not exceed the discrepancy encoded in $\mathcal{A}_\mu^{\rm BSM}$, we obtain a $95\%$~C.L. bound on $C_{\ell 1}^{ds}$ of $[-0.1,3.6]\times10^{-6}$, which translates into a $95\%$~C.L. upper bound on $c_{uW}^{11}$,
\begin{equation}
    |c_{uW}^{11}|<0.5\left(\Lambda /\tev\right)^2\;.	
\end{equation}
This bound, the strongest of the ones that we derive in this section, is weaker than the projected collider bounds for FCC-hh presented in Tab.~\ref{tab:bounds_summary} and, in particular, at best of the same order as our estimates for LHC Run 2.
Turning now to the dipole operator $\mathcal{Q}_{7\gamma}^{(\prime)}$, it  can induce $d_i\rightarrow d_j\gamma$ transition. The magnetic and electric dipole WET operators are commonly defined in the basis~\cite{Mertens:2011ts} as follows,
\begin{equation}
   Q_{\gamma}^{\pm}=\frac{Q_de}{16\pi^2}(\bar{s}_L\sigma^{\mu\nu}d_R\pm \bar{s}_R\sigma^{\mu\nu}d_L)F_{\mu\nu}\;,
\end{equation}
where we can relate $C_{7\gamma}^{(\prime)}$ to the Wilson coefficients of the new operators via
\begin{equation}
C_{7\gamma}^{21}=\frac{Q_d\left(C_\gamma^++C_\gamma^-\right)}{16\pi^2 G_F m_d}\;,\quad  C_{7\gamma}^{\prime 21}=\frac{Q_d\left(C_\gamma^+-C_\gamma^-\right)}{16\pi^2 G_F m_s}\;.
\end{equation}
In the SM, the estimated value of the real component of $C_{\gamma}^\pm$, denoted as $\left|{\rm Re}\, C_{\gamma}^\pm\right|^{\rm SM} $, is approximately $ 0.06\, G_F\, m_K$, and the experimental constraints on $C_{\gamma}^\pm$ are derived from the processes $K^+\rightarrow\pi^+\pi^0\gamma$ and $K^0\rightarrow\gamma\gamma$~\cite{Mertens:2011ts}. The $s\rightarrow d\gamma$ contributions consist of an electric ($\propto C_\gamma^-$) and magnetic component $(\propto C_\gamma^+)$. Among them, the electric amplitude is more precisely measured and leads to the constraint, $|{\rm Re}\, C_\gamma^-|<0.1\,G_F\, m_K$. 

For $K_S\rightarrow \gamma(k_1,\mu)\gamma(k_2,\nu)$ decay amplitudes, the photons produced in this decay have parallel polarization, $\mathcal{A}(K^0\rightarrow (\gamma\gamma)_\parallel)\times\left(k_1^\nu k_2^\mu-k_1\cdot k_2 g^{\mu\nu}\right)$, while the photon produced by $K_L\rightarrow\gamma\gamma$ have perpendicular polarization, $\mathcal{A}(K^0\rightarrow (\gamma\gamma)_\perp)\times i\varepsilon_{\mu\nu\rho\sigma}k_{1}^\rho k_2^\sigma$ . In this notation, the decay width is given by $\Gamma(K^0\rightarrow\gamma\gamma)=m_K^3|\mathcal{A}|^2/64\pi$. Parametrizing the amplitudes as~\cite{Mertens:2011ts},
\bea
\mathcal{A}(K^0\rightarrow (\gamma\gamma)_\parallel)=\frac{A_{\gamma\gamma}^\parallel}{\sqrt2}\times(\alpha_{\rm em} G_F m_K)\;, \quad \mathcal{A}(K^0\rightarrow (\gamma\gamma)_\perp)=\frac{A_{\gamma\gamma}^\perp}{\sqrt2}\times(\alpha_{\rm em} G_F m_K)\;,
\eea
we can fix $|A_{\gamma\gamma}^\parallel|^{\rm exp}=0.191$ and $|A_{\gamma\gamma}^\perp|^{\rm exp}=0.115$ from the $K_{L,S}\rightarrow\gamma\gamma$ decay rates~\cite{Workman:2022ynf}. 
The dipole operator $Q_\gamma^{\pm}$ contribute to the kaon diphoton decay amplitude~\cite{Mertens:2011ts} as
\begin{equation}
    \Delta A_{\gamma\gamma}^{\parallel,\perp}=\frac{2F_\pi B_T'C_{\gamma}^{-,+}}{9\pi G_F m_K^2}\;,
\end{equation}
where  $F_\pi=92.4\,\mev$ and $B_T'=2.67(17)$ is extracted from the lattice estimation~\cite{Buividovich:2009bh}. The SM prediction,  $|A_{\gamma\gamma}^\parallel|^{\rm SM}=0.166$ and $|A_{\gamma\gamma}^\perp|^{\rm SM}=0.126$ provides constraints $|{\rm Re}\, C_\gamma^-|<0.7\,G_F\, m_K$, and $ |{\rm Re}\, C_\gamma^+|< 0.3\,G_F\, m_K$. This turns into the following bound
\begin{equation}
	|c_{uW}^{11}|<40.3\left(\Lambda /\tev\right)^2\;.
\end{equation}
 from the $K^0\rightarrow\gamma\gamma$ channel.

\subsection{FCCC}

Beyond the aforementioned loop-level FCNC, our flavor assumptions lead to tree-level FCCC. Some of those, including the pion decay $\pi^-\rightarrow e^-\bar{\nu}_e\gamma$ on which we focus below, can be worked out by treating the $W$ boson field strength in the EW dipole of Eq.~\eqref{eq:CqW} as an external source that extends the QCD Lagrangian, and that ought to be integrated out afterwards in $\chi$PT. In particular, the effect of other tree-level operators such as 4-quark operators is subleading.

 The decay $\pi^-\rightarrow e^-\bar{\nu}_e\gamma$ decay requires an operator with a $W$ boson and a photon. Combining the $W$ boson field strength and the SMEFT WC $c_{uW}$ in Eq.~\eqref{eq:CqW} into a chiral tensor spurion $t_{\mu\nu}$, we follow Refs.~\cite{Cata:2007ns,Mateu:2007tr} and match to the lowest-order operator involving a photon field strength in $\chi$PT,
\begin{equation}\label{eq:ChptTensor}
	\mathcal{L}^{\chi {\rm PT}}\supset \frac{B_0 F_\pi^2}{M_\rho^2}\tr\left( t_+^{\mu\nu}f_{+\mu\nu}\right)\;,
\end{equation}
where $B_0F_\pi^2=-\langle\bar{q}q\rangle=-[242(15) \,{\rm MeV}]^3$~\cite{Jamin:2002ev}, $M_\rho=0.775\,\gev$ is the $\rho$ meson mass, and the definitions of $t_+^{\mu\nu}$ and $f_{+\mu\nu}$ are given in App.~\ref{app:tensorsource}. Expanding to leading order in $F_\pi$ one finds,
\bea\label{eq:pidecay}
	\mathcal{L}^{\chi {\rm PT}}_1\supset \frac{i 2\, e\, B_0\, F_\pi G_F v_hV_{ud}\,c_{uW}^{11}}{3g_2M_\rho^2 \Lambda^2} \partial_\mu\pi^+J_\nu^-\left(F^{\mu\nu}+i\tilde{F}^{\mu\nu}\right)+{\rm h.c.}\;,
\eea
where $J_\mu^-=\bar{\nu}\gamma^\mu P_L e$ is the weak current and $\tilde{F}_{\mu\nu}=\epsilon_{\mu\nu\rho\sigma}\,F^{\rho\sigma}/2$. This implies that the quark dipole can induce additional contributions to the form factors $F_V$ and $F_A$ describing the $\pi^-\rightarrow e^-\bar{\nu}_e\gamma$ decay amplitude $\mathcal{M}_{\rm SD}$~\cite{Belyaev:1991gs,Bolotov:1990yq},
\begin{equation}\label{eq:pidecaysm}
	\mathcal{M}_{\rm SD}=-\frac{e\,G_{\rm F}V_{ud}}{\sqrt2 m_\pi}\epsilon^{*\mu}\left[F_V\epsilon_{\mu\nu\sigma\tau}\, p^\sigma q^\tau +iF_A(g_{\mu\nu}\, p\cdot q - p_\nu q_\mu )\right]\bar{u}\gamma^\nu(1+\gamma^5)v \ .
\end{equation}
From Eq.~\eqref{eq:pidecay} and ~\eqref{eq:pidecaysm}, we get,
\begin{equation}
	\Delta F_V=\Delta F_A=\frac{4\sqrt{2}\, c_{uW}^{11}\, B_0 F_\pi v_hm_\pi}{3\, g_2\, M_\rho^2 \Lambda^2}\;.
\end{equation}
The theoretical values of the pion form factor is related to the conserved vector current to the $\pi^0\rightarrow \gamma\gamma$ decay width~\cite{muller1963connection, Unterdorfer:2008zz}, 
\bea
F_V^{\rm SM}&=\alpha_{\rm em}^{-1}\sqrt{2\,\Gamma_{\pi^0\rightarrow\gamma\gamma}/(\pi m_{\pi^0})}=0.0262(5)\\ F_A^{\rm SM}&=4\sqrt2\left(L_9+L_{10}\right)m_{\pi^+}/F_\pi=0.0106(36) \ .
\eea
In this context, $L_9\simeq 6.49\times 10^{-3}$ and $L_{10}\simeq -5.10\times 10^{-3}$ are the coefficients within the general $\chi$PT of the order of $\mathcal{O}(p^4)$ as outlined in~\cite{Gasser:1984gg, Bijnens:1992en}. The value of $L_9$ can be determined from the charge radius of the pion. Recent analysis of the CERN SPS experiment~\cite{NA7:1986vav} and the new data by the  JLAB-$\pi$ Collaboration~\cite{JeffersonLab:2008jve} imply that $L_9$ differs by $\simeq 1.6$ standard deviations~\cite{Simula:2023ujs} from the current result. We thus expect to improve the theoretical prediction of $F_A^{\text{SM}}$ in the near future.
The experimental values of the form factors are measured by the CsI crystal calorimeter of PIBETA group~\cite{Bychkov:2008ws}, $F_V^{\rm exp}=0.0258(17)$, $F_A^{\rm exp}=0.0117(17)$. Assuming that the dipole contribution to $F_{V, A}$ 
can not exceed the discrepancy between the SM prediction and the measured value we can bound the dipole WC as,
\begin{equation}
	|c_{uW}^{11}|<13\left(\Lambda /\tev\right)^2\;.
\end{equation}

\subsection{Quark Mass Naturalness}\label{sec:naturalnessQuarkMasses}

Finally, we investigate naturalness constraints on the quark masses. As we are introducing additional spurions beyond the quark Yukawas with sizable entries along the light generations, radiative contributions to the low-energy quark masses potentially lead to an up-quark hierarchy problem.

The one-loop diagram that renormalizes the up-quark mass with one dipole insertion gives,
\bea
    \frac{\delta m_u}{m_u}=\frac{3c_{uW}^{11} m_W}{2m_u\Lambda^2}\int_0^E \frac{d^4 q}{(2\pi)^4}\frac{q_\mu\sigma^{\mu\nu}\slashed{q}\gamma_\nu}{q^2(q^2-m_W^2)}
    \sim\frac{3c_{uW}^{11} m_W E^2}{16\pi^2 m_u\Lambda^2}\;,
\eea
where we considered a hard cutoff $E$ to extract the quadratic divergence, which we take as a proxy for possible threshold corrections at scales $\sim\Lambda$. If we require one percent tuning, $|\delta m_u/m_u|<100$, the quark Yukawa naturalness provides an upper bound on the Wilson coefficient,
\begin{equation}
    |c_{uW}^{11}|\lesssim 0.1
    \;.
\end{equation}
In appropriate UV models, threshold corrections to the small up-quark mass can be controlled employing symmetry. For instance, Ref.~\cite{Voloshin:1987qy} argues that a suppressed neutrino mass accompanied by a sizeable magnetic dipole could result from an $SU(4)$ symmetry in the lepton sector. A similar symmetry in the quark sector could suppress the light-quark Yukawa couplings, but we leave the development of a precise model for future work.

\section{A Renormalizable UV Model }\label{sec:simpmodel}
In this section, we explore fully renormalizable theories that can give rise to $c_{uW}$ as in Eq.~\eqref{eq:cuvihmfv}, and to no other dangerous flavorful SMEFT operators. As IHMFV is a subclass of alignment as we defined it, we take this model as proof of principle that either of them can be obtained from weakly-coupled UV completions.

To this end, we introduce a set of heavy vector-like fermions $F$, $X$ and a complex scalar $S$. The fermions come in three generations, and the SM charges of the new fields are listed in Tab.~\ref{tab:FSX}.
\begin{table}
\begin{center}
  \begin{tabular}{ | c | c | c | c | c | c | c | c|}
  \hline 
  BSM Fields  & $F$& $S$ & $X$  \\ 
   \hline
 $G_{\rm SM}$  & $(\mathbf{3},\mathbf{2},\frac16)$ & $(\mathbf{1},\mathbf{1}, 0)$ & $(\mathbf{3},\mathbf{2},\frac16)$\\
   \hline
  \end{tabular}
  \end{center}
\caption{Standard Model gauge charges of the heavy BSM fields of our renormalizable UV model.}
\label{tab:FSX}
\end{table}
As we are aiming at deriving IHMFV, we need to specify the (spurious) flavor transformations of all building blocks (fields and couplings) of the theory. IHMFV arising from regular MFV in the UV, we assume that the flavor non-universal interactions depend on a single spurion $Y_u$. Focusing on the active subset $G_F^q\equiv SU(3)_Q\times SU(3)_u$ of the flavor group, we assume the transformations in Tab.~\ref{tab:global_charges_UV}. 
\begin{table}
\begin{center}
  \begin{tabular}{ | c | c | c | c | c | c | c | c || c|}
  \hline 
  Names  &$Q_L$&$u_R$& $\mathbb{P}_LF$ &$\mathbb{P}_RF$& $\mathbb{P}_LX$& $\mathbb{P}_RX$& $S$ 
& $Y_u$ \\ 
   \hline
 $G_{F}^q$ &$(\mathbf{3},\mathbf{1})$&$(\mathbf{1},\mathbf{3})$ & 
 \multicolumn{2}{c|}{$(\mathbf{3},\mathbf{1})$}& $(\mathbf{1},\mathbf{3})$&$(\mathbf{3}, \mathbf{1})$&$(\mathbf{1},\mathbf{1})$
&$(\mathbf{3}, \mathbf{\bar{3}})$\\
   \hline
      $U(1)_M$ & \multicolumn{2}{c|}{$0$}& \multicolumn{2}{c|}{$q$}&\multicolumn{2}{c|}{$-q$}&\multicolumn{2}{c|}{$0$}\\
   \hline
  \end{tabular}
  \end{center}
 \caption{Global symmetry charges of the relevant fields in our renormalizable UV model. Among them $Y_u$ functions as a spurion, whereas the remaining fields are dynamical.}
 \label{tab:global_charges_UV}
 \end{table}
We also indicate there the charges under a global unbroken $U(1)_M$ symmetry. No new field is introduced in the down sector, and $Y_d$ is the only spurion with down-type charges. Therefore, IHMFV will only be active in the up sector -- only inverse powers of $Y_u$ will be present in the SMEFT Wilson coefficients, while $Y_d$ will always appear as in regular MFV.
The most general Lagrangian respecting all (spurious) symmetries reads
\bea\label{eq:simpUV}
\mathcal{L}=&\bar{Q}_L \lambda_{q} F S+\bar{F}\lambda_F\mathbb{P}_RS^\dagger X+\bar{X}\lambda_{u}u_R\tH+\bar{X} M Y_u^\dagger\mathbb{P}_R X+\bar{Q}_L m X\\
&+\bar{X}\xi Y_u^\dagger \mathbb{P}_R F S+\bar{Q}_LY_uu_R \tH +\bar{F} M_F\mathbb{P}_RF+{\rm h.c.}+m_S^2|S|^2 \ .
\eea 
$m$ mixes $Q_L$ with X, but this mixing can be removed by a field redefinition.\footnote{The mass mixing term in Eq.~\eqref{eq:simpUV} can be removed by redefining $X_L$ and $Q_L$ via,
\begin{equation*}
    X_L^{\prime\dagger}=\left(X_L^\dagger Y_u^\dagger +\alpha Q_L^\dagger\right)\left(Y_u^\dagger Z\right)^{-1}\;,\quad Q_L^{\prime\dagger}=\left(-\alpha X_L^\dagger + Q_L^\dagger Y_u\right)\left(Z Y_u\right)^{-1}\;,
\end{equation*}
where $ \alpha=m/M$ and $ Z=\sqrt{1+\alpha^2 X_u^{-1} }$, so that the new Yukawa matrix reads
\begin{equation*}
   Y_u'=(1-\alpha\lambda_u X_u^{-1})Z^{-1}Y_u\;.
\end{equation*}} Therefore, in what follows we consider $m=0$.
Most masses and Yukawa couplings above transform in a singlet or adjoint representation of the flavor group: $\lambda_q, \lambda_F,m,M_F$ transform as $(\mathbf{1} \oplus \mathbf{8},\mathbf{1})$, while $\lambda_u,M,\xi$ transform as $(\mathbf{1}, \mathbf{1} \oplus \mathbf{8})$. Therefore, they admit an expansion in terms of the only flavorful building blocks, $Y_{u,d}$. Assuming only positive powers of those spurions, we have
\begin{equation}\label{eq:lamqmfv}
    \lambda_{q} \propto \mathbb{1}+a_1 X_u+a_2 X_d+\cdots\;,\quad \lambda_u\sim \mathbb{1}+b_2 Y_u^\dagger Y_u+\cdots \;,
\end{equation}
where $X_{u,d}\equiv Y_{u,d}^{\phantom{\dagger}} Y_{u,d}^{\dagger}$, as defined in Sec.~\ref{sec:EFT_framework}.
This expansion leads to flavor-changing matrix elements suppressed by the normal MFV mechanism. We stress that renormalization group running will slightly misalign the flavor structures of the various couplings, so that they will not be captured by a single spurion $Y_u$ at all scales. Therefore, the Lagrangian in Eq.~\eqref{eq:simpUV} is meant to hold at a high scale, for instance that of spontaneous breaking of the flavor group where the Yukawas are generated, if such a mechanism occurs. Nevertheless, the hierarchies inherited at the high scale, which are the main focus of this work, will not be drastically affected by that running.

Integrating out the BSM fields $F$, $S$ and $X$ at one loop we get,
 \begin{equation}\label{eq:CuV}	
 \frac{c_{uW}}{\Lambda^2}=\frac{g_2\lambda_q\lambda_F M_F \hat{M}_X^{-1}\lambda_u}{128\pi^2m_S^2 }\mathcal{F}_2^C(r) \ ,
 \end{equation}
where $\hat{M}_X\equiv Y_u^\dagger M$ is the mass matrix of $X$ and the loop function $\mathcal{F}_2^C$ is given by
\begin{equation}
\mathcal{F}_2^C\left(r\right) = \frac{3}{2}\frac{[3-4r+r^2+2 \ln(r)]}{(r-1)^3}, \quad\text{with } r=\frac{(y_2\, M)^2}{m_S^2} \ .
\end{equation}
The SMEFT cut-off scale is set by the lightest BSM particles, $\Lambda\sim \text{min}\left(M_S,M_F,y_u M\right)$. In Eq.~\eqref{eq:CuV}, $\hat{M}_X$ is the only hierarchical matrix in flavor space, so the flavor structure of $c_{uW}$ is dictated by the inverse Yukawa, $c_{uW}\propto \widetilde{Y}_u$.
This realizes IHMFV. 
Due to the MFV expansion of $\lambda_{q,u}$, the off-diagonal components of $c_{uW}$ in the up basis is suppressed by $\lamfc$,
\begin{equation}
    c_{uW}^{12}\sim \lambda^{11}c_{uW}^{11}\;.
\end{equation}
 We stress again that the fact that the off-diagonal elements of $c_{uW}$ are proportional to entries of $X_d$ in the up-basis, and not of $\tilde X_d$ which would yield a much weaker suppression, is due to the fact that we did not introduce any BSM field whose mass arises from $Y_d$. 
 
 As discussed in Sec.~\ref{sec:IHMFV}, in IHMFV large flavor-changing effects can be mediated by other SMEFT operators than the dipole, in particular by the LL-type 4-quark operators. However, those turn out to be flavor-diagonal in the model presented here. Other operators which do not feature left-handed quarks turn out to be flavor diagonal in the mass basis, as they should given the underlying aligned IHMFV structure.
 A detailed discussion can be found in App. ~\ref{app:Oeff}, where we
show the dimension-6 $Q_LQ_L$-type SMEFT operators induced by the model in Eq.~\eqref{eq:simpUV}, up to one-loop level. In this appendix, we also briefly assess the collider phenomenology associated to these additional operators. A complete list of dimension-6 SMEFT operators generated by \texttt{Matchmakereft}~\cite{Carmona:2021xtq} is available upon request.

\section{Conclusion and Discussion}\label{sec:conclusion}
 
SMEFT provides a convenient and consistent description of physics beyond the SM when the new particles are heavy enough to not be produced on-shell at a given experiment.
The effects of different UV models are encoded in the Wilson Coefficients that weigh the SMEFT higher-dimensional operators. Given that the vast majority of the free coefficients in the SMEFT is associated with the existence of three generations, {\it i.e.} flavor physics, and given that the bounds arising from searches for non-standard flavor violation vastly dominate other ones for random WCs, it is important to deeply investigate the flavor structure of the SMEFT. This has been done in the past from a variety of perspectives, most notably from that of specific UV models and from that of flavor symmetries.

In the realm of flavor symmetries, the most common framework is Minimal Flavor Violation, where one assumes that the only flavor-breaking spurions that one can use to build SMEFT WCs are the SM Yukawa couplings. This strongly alleviates FCNC constraints on the scale of new physics. However the MFV paradigm only represents a subset of the possible EFTs that evade flavor bounds. In this work, we scrutinized a flavor structure in the SMEFT which has been dubbed flavor alignment in previous literature. This structure is similar to MFV in that flavor violation is fully dictated by the SM CKM matrix, while it radically differs in that WCs do not follow the same hierarchies as quark masses and couplings to the Higgs boson.
In short, light generations are allowed to couple to large flavor-breaking spurions which are only misaligned from quark masses by the CKM matrix. We also investigated a more restrictive assumption, which we call \textit{inverse hierarchy MFV}, in which the new aligned spurions are proportional to the inverse of the Yukawa matrices.

As it stands, those assumptions are not as efficient as MFV to weaken FCNC bounds. However, this depends on the set of SMEFT operators under consideration. We focused on the up-type quark electroweak dipole operator $\mathcal{O}_{uW}$, for which we demonstrated that the bounds from flavor violation are rather weak, even for dipoles of the light quark generations. 
Instead, we showed that those dipoles can be equally well constrained from current LHC data, and will be much better probed at future hadron colliders such as HL-LHC and FCC-hh.

\begin{figure}
\centering
\includegraphics[width=0.74\linewidth]{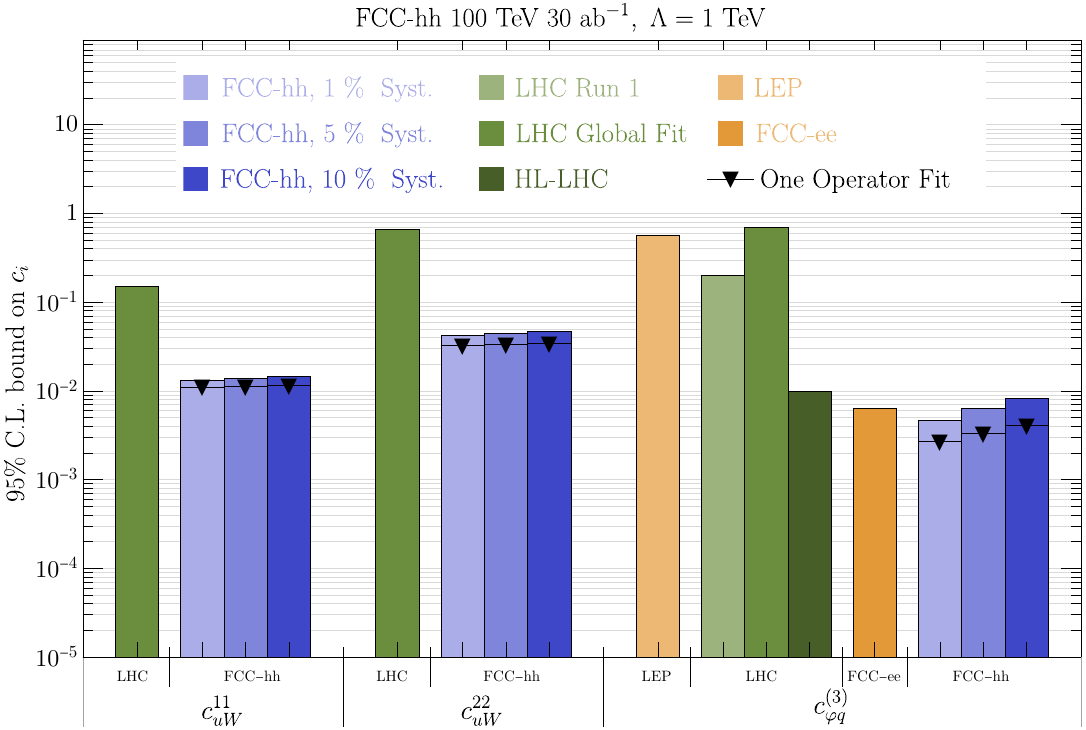}
\caption{
$95\%$ C.L. bounds on $c_{uW}^{11}$, $c_{uW}^{22}$, and  $c_{\varphi q}^{(3)}$. We show in blue our bounds from a one-operator fit of $Wh \to \ell \nu \gamma \gamma$ at FCC-hh with 30 ab$^{-1}$ for different systematic uncertainties. The filled bars give the bounds after profiling over the other WCs in the analysis, while the horizontal black line with a triangle indicates the bounds from one-operator fits. In green for $c_{uW}^{11}$ and $c_{uW}^{22}$, the bounds are taken from the HighPT fit~\cite{Allwicher:2022gkm}. For $c_{\varphi q}^{(3)}$, the lighter and darker green bounds are from LHC and HL-LHC using leptonic $WZ$~\cite{Franceschini:2017xkh}, and the bounds in medium green is from a global fit~\cite{Ellis:2018gqa}. In lighter and darker orange for $c_{\varphi q}^{(3)}$, the bounds are from LEP~\cite{Franceschini:2017xkh} and FCC-ee global fit~\cite{DeBlas:2019qco}.
}
\label{fig:pThv3}
\end{figure}

More precisely, this can be done by looking at the high-energy tails in $Wh$ production at hadron colliders. Indeed, the $pp\rightarrow Wh$ process is dominated by the light quark operators allowed by our flavor assumptions. We showed how a simple analysis at FCC-hh could provide competitive bounds when the final state is $\ell\nu\gamma\gamma$. 
A simple scaling, based on the comparison with previous analyses, showed that the same process, but with the Higgs decaying to $b\bar b$, would yield relevant bounds at LHC Run 2, Run 3 and HL-LHC, even in comparison with current and projected bounds from Drell-Yan high-energy tails.
In Fig.~\ref{fig:pThv3}, we summarize our projected bounds for FCC-hh and other current or projected bounds. We only show the general flavor scenario since the flavor-universal case gives results very similar to the ones for the first-generation quarks. This plot shows the potential of $Wh$ production as a probe of EW dipoles in the flavor scenario presented in this work.

Finally, we presented a UV model that realizes our flavor and operator assumptions. Its UV flavor structure follows the regular MFV assumption, and it features heavy massive fields that transform chirally under the flavor group, so that the induced low-energy EFT follows the inverse hierarchy MFV. An appropriate choice of the UV fields and their charges ensures that only the dipole operator violates flavor at the leading order. 

This work represents a first step towards the exploration of exotic but allowed flavor structures in SMEFT and other similar EFTs and hence the paths for future exploration abound. How our flavor assumptions can be extended to operators of dimension bigger than six is one of them, for which our UV-completed model might provide a solid stepping stone. The phenomenology of such higher-dimensional operators is equally promising for future research. Although our study includes a comprehensive overview of experimental probes of the aligned flavor scenario, a deeper exploration could provide new experimental probes both from collider and low-energy experiments. The precise study of RGE effects on our flavor assumptions is an interesting avenue, as well as a detailed study of all UV models that could realize either flavor alignment or, more specifically, IHMFV.

\section*{Acknowledgments}
We are particularly grateful to Christophe Grojean for his collaboration during the early stages of the project.
We thank Fady Bishara, David Friday, Guilherme Guedes, Thomas McKelvey, Peter Stangl, Luiz Vale Silva, Eleni Vryonidou, Ludovico Vittorio and Jure Zupan for discussions.  This work is supported by the Deutsche Forschungsgemeinschaft under Germany’s Excellence Strategy EXC 2121 “Quantum Universe” -- 390833306, as well as by the grant 491245950. This project also has received funding from the European Union’s Horizon Europe research and innovation programme under the Marie Skłodowska-Curie Staff Exchange grant agreement No 101086085 - ASYMMETRY. C.Y.Y. is supported in part by the Grants No. NSFC-11975130, No. NSFC-12035008, No. NSFC-12047533, the National Key Research and Development Program of China under Grant No. 2017YFA0402200, the China Postdoctoral Science Foundation under Grant No. 2018M641621 and the Helmholtz-OCPC International Postdoctoral Exchange Fellowship Program. Q.B. received support from the DOE Early Career Grant DESC0019225, and is currently supported by a starting grant of the French Ministère de l'Enseignement supérieur et de la Recherche.
A. N. R. has been supported by the European Research Council (ERC)
under the European Union’s Horizon 2020 research and innovation programme (Grant
agreement No. 949451) and a Royal Society University Research Fellowship through grant URF/R1/201553 during the completion of this project. A. N. R. acknowledges support from the COMETA COST Action CA22130. D.L. acknowledges funding from the French Programme d’investissements d’avenir through the Enigmass Labex.


\appendix

\section{$Wh$ Analysis}
\label{app:Pheno_analysis}

\subsection{Collider Event Simulation}
\label{app:Pheno_analysis_simulation}

The dependence of the signal events on the EW dipole operators was computed via Monte Carlo simulations with \texttt{MadGraph5\_aMC@NLO v.3.3.2}~\cite{Alwall:2014hca} and the UFO model \texttt{SMEFTsim v.3.0.2}~\cite{Brivio:2020onw}. The simulations were performed at LO and NLO QCD and EW corrections were accounted for via $p_{T}^h$-dependent k-factors, listed in Tab.~\ref{tab:k-Factors}. The SM signal number of events as well as its dependence on $c_{\varphi q}^{(3)}$ and the background number of events were extracted from Ref.~\cite{Bishara:2020vix}. There, the signal and the main background, $W\gamma\gamma$, were simulated with merged $0+1$ jet samples in order to account for the leading NLO QCD corrections, while the NLO EW corrections were applied via k-factors. The sub-leading backgrounds $Wj\gamma$ and $Wjj$ were simulated at LO QCD, and a flat jet-to-photon mistagging rate of $10^{-3}$ was assumed.

\begin{table}[htb!]
	\centering
	\begin{tabular}{c|c|c}
		\toprule
		$p_{T}^{h}$ bin & $k_{\rm QCD}$ & $k_{\rm EW}$ \tabularnewline
		\midrule
		$[200-400)$ & $0.286$ & $-0.08$ \tabularnewline
		$[400-600)$ & $0.36$ & $-0.15$ \tabularnewline
		$[600-800)$ & $0.42$ & $-0.21$\tabularnewline
		$[800-1000)$ & $0.41$ & $-0.27$\tabularnewline
		$[1000-\infty)$ & $0.41$ & $-0.40$ \tabularnewline
		\bottomrule
	\end{tabular}
	\caption{k-Factors for $pp\to \ell\nu h$ in the SM. They are defined such that $\sigma^{\rm NLO}/\sigma^{\rm LO} =1+k_{\rm QCD}+k_{\rm EW}$.}
	\label{tab:k-Factors}
\end{table}

\subsection{Event Selection}
\label{app:Pheno_analysis_cuts}

The analysis strategy is the same as the one adopted in Ref.~\cite{Bishara:2020vix}. The sought-after final state consists of 2 photons that reconstruct the Higgs boson,  ensured by a cut on the photon pair invariant mass $m_{\gamma\gamma}\in[120,130]$~GeV, a hard charged lepton of the first 2 generations and at least $100$~GeV of missing transverse energy, $\slashed{E}_T$. To increase the signal fraction in the selected events, we further impose cuts on the angular distance of the diphoton pair and on the maximum $p_{T}$ of the $Wh$ system, as detailed in Tab.~\ref{tab:app_selection_cuts}. An analysis of the effectiveness of these cuts can be found in Ref.~\cite{Bishara:2020vix}.

\begin{table}[h!]
\begin{centering}
\begin{tabular}{c|c}
\multicolumn{2}{c}{Selection cuts}\\
\hline
$p_{T,\text{min}}^{\ell}$ & $30$~GeV\\
$p_{T,\text{min}}^{\gamma}$ & $50$~GeV\\
$\slashed{E}_{T,\text{min}}$ & $100$~GeV\\
$m_{\gamma\gamma}$ & $[120,130]$~GeV\\
\hline
$\Delta R_{\text{max}}^{\gamma\gamma}$ & $\lbrace 1.3,0.9,0.75,0.6,0.6 \rbrace$\\
$p_{T,\text{max}}^{Wh}$ & $\lbrace 300,500,700,900,900 \rbrace$~GeV
\end{tabular}
\par\end{centering}
\caption{Cuts used to select Monte Carlo events. The last two rows correspond to cuts that depend on the $p_{T}^{h}$ bin and the entries in the list correspond to each $p_T^h$ bin, as defined in Tab.~\ref{tab:double_bin}.}
\label{tab:app_selection_cuts}
\end{table}

\subsection{Signal and Background Cross Section}
\label{app:Pheno_analysis_xs}

In Tab.~\ref{tab:sigma_full}, we present the estimated number of events in each bin for signal and background at FCC-hh with $\mathcal{L}=30$~ab$^{-1}$. For the signal, we report the number of events only as a function of the dipole WCs $c_{uW}^{11}$ and $c_{uW}^{22}$. Its dependence on the WCs $c_{\varphi q}^{(3)}$, $c_{\varphi W}$ and $c_{\varphi \tilde W}$ can be found in Ref.~\cite{Bishara:2020vix}.

\begin{table}[h!!]
\begin{centering}
\setlength{\extrarowheight}{0mm}%
\scalebox{.86}{
\begin{tabular}{c|c|c|c}
\toprule
\rule[-.5em]{0pt}{.5em}
\multirow{2}{*}{$p_T^{h}$ bin} &\multirow{2}{*}{$\phi_{W}$ bin}& \multicolumn{2}{c}{Number of expected events}\tabularnewline
\cline{3-4} &  & Signal & Background \tabularnewline
\hline 
\multirow{2}{*}{\rule{0pt}{3em}$[200-400]$\,GeV} & $[-\pi,0]$ &
  $\begin{aligned}\rule{0pt}{1.15em}
1310\, +\,& 26700 \left(c_{u W}^{11}\right)^2+5300 \left(c_{u W}^{22}\right)^2\\
+\,& 10380 c_{\varphi q}^{(3)}+25700 \left(c_{\varphi q}^{(3)}\right)^{2}
\rule[-.5em]{0pt}{1.em}
\end{aligned}
$ & $830$\tabularnewline
\cline{2-4} 
  & $[0,\pi]$ & $\begin{aligned}\rule{0pt}{1.15em}
1310 \, +\,& 25200 \left(c_{u W}^{11}\right)^2+5380 \left(c_{u W}^{22}\right)^2\\
+\,& 10480 c_{\varphi q}^{(3)}+27000 \left(c_{\varphi q}^{(3)}\right)^{2}
                  \rule[-.5em]{0pt}{1.em}
\end{aligned}
$ & $960$\tabularnewline
\hline 
    \multirow{2}{*}{\rule{0pt}{3em}$[400-600]$\,GeV} & $[-\pi,0]$ & $\begin{aligned}\rule{0pt}{1.15em}
284 \, +\,& 36500 \left(c_{u W}^{11}\right)^2+6300 \left(c_{u W}^{22}\right)^2\\
+\,& 5820 c_{\varphi q}^{(3)}+35800 \left(c_{\varphi q}^{(3)}\right)^{2}     \rule[-.5em]{0pt}{1.em}
\end{aligned}
$ & $119$\tabularnewline
\cline{2-4} 
  & $[0,\pi]$ & $\begin{aligned}\rule{0pt}{1.15em}
283 \, +\,& 36600 \left(c_{u W}^{11}\right)^2+6290 \left(c_{u W}^{22}\right)^2\\
+\,& 5860 c_{\varphi q}^{(3)}+36000 \left(c_{\varphi q}^{(3)}\right)^{2}   \rule[-.5em]{0pt}{1.em}
\end{aligned}
$ & $129$\tabularnewline
\hline 
    \multirow{2}{*}{\rule{0pt}{3em}$[600-800]$\,GeV} & $[-\pi,0]$ & $\begin{aligned}\rule{0pt}{1.15em}
70 \, +\,& 37300 \left(c_{u W}^{11}\right)^2+5610 \left(c_{u W}^{22}\right)^2\\
+\,& 2760 c_{\varphi q}^{(3)}+33500 \left(c_{\varphi q}^{(3)}\right)^{2}\rule[-.5em]{0pt}{1.em}
\end{aligned}
$ & $21$\tabularnewline
\cline{2-4} 
  & $[0,\pi]$ & $\begin{aligned}\rule{0pt}{1.15em}
70 \, +\,& 36700 \left(c_{u W}^{11}\right)^2+5590 \left(c_{u W}^{22}\right)^2\\
+\,& 2850 c_{\varphi q}^{(3)}+33800 \left(c_{\varphi q}^{(3)}\right)^{2}   \rule[-.5em]{0pt}{1.em}
\end{aligned}
$ & $22$ \tabularnewline
\hline 
    \multirow{2}{*}{\rule{0pt}{3em}$[800-1000]$\,GeV} & $[-\pi,0]$ & $\begin{aligned}\rule{0pt}{1.15em}
15 \, +\,& 20300 \left(c_{u W}^{11}\right)^2+2700 \left(c_{u W}^{22}\right)^2\\
+\,& 947 c_{\varphi q}^{(3)}+17900 \left(c_{\varphi q}^{(3)}\right)^{2}\rule[-.5em]{0pt}{1.em}
\end{aligned}
$ & $3$ \tabularnewline
\cline{2-4} 
  & $[0,\pi]$ & $\begin{aligned}\rule{0pt}{1.15em}
15 \, +\,& 20400 \left(c_{u W}^{11}\right)^2+2800 \left(c_{u W}^{22}\right)^2\\
+\,& 947 c_{\varphi q}^{(3)}+18200 \left(c_{\varphi q}^{(3)}\right)^{2} \rule[-.5em]{0pt}{1.em}
\end{aligned}
$ & $5$ \tabularnewline
\hline 
    \multirow{2}{*}{\rule{0pt}{3em}$[1000-\infty]$\,GeV} & $[-\pi,0]$ & $\begin{aligned}\rule{0pt}{1.15em}
4 \, +\,& 21900 \left(c_{u W}^{11}\right)^2+2230 \left(c_{u W}^{22}\right)^2\\
+\,& 426 c_{\varphi q}^{(3)}+16400 \left(c_{\varphi q}^{(3)}\right)^{2} \rule[-.5em]{0pt}{1.em}
\end{aligned}
$ & $2$ \tabularnewline
\cline{2-4} 
  & $[0,\pi]$ & $\begin{aligned}\rule{0pt}{1.15em}
4 \, +\,& 21600 \left(c_{u W}^{11}\right)^2+2210 \left(c_{u W}^{22}\right)^2\\
+\,& 428 c_{\varphi q}^{(3)}+16600 \left(c_{\varphi q}^{(3)}\right)^{2}  \rule[-.5em]{0pt}{1.em}
\end{aligned}$
 & $1$ \tabularnewline
\bottomrule
\end{tabular}
}
\par\end{centering}
\caption{Number of expected signal and background events at FCC-hh with $30\,{\rm ab}^{-1}$. For the signal, it is given as a function of the Wilson coefficients (with $\Lambda = 1 \, \text{TeV}$).  Notice that the coefficients have errors of order $\it few$ percent due to statistical fluctuations. The contribution of $Wjj$ to the background events is neglected.
}
\label{tab:sigma_full}
\end{table}

\begin{figure}
        \centering
        \includegraphics[width=0.74\linewidth]{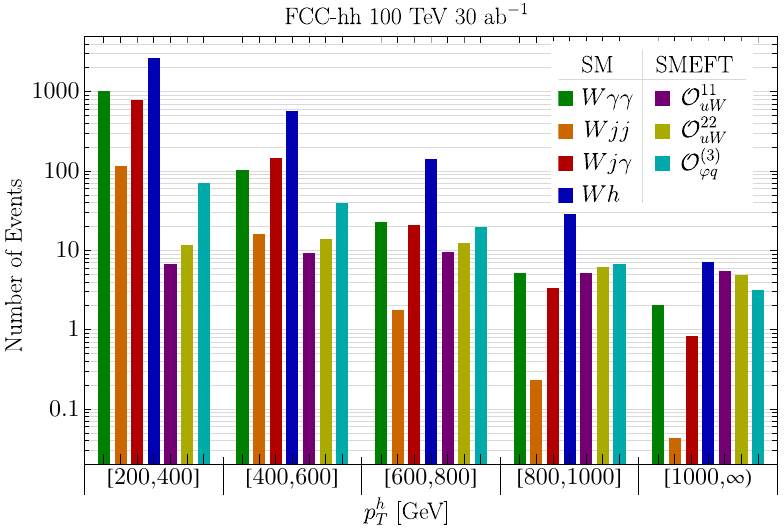}
        \caption{Number of SM and SMEFT events per $p_T^{h}$ bin after selection cuts for the signal and backgrounds at the FCC-hh assuming 30 ab$^{-1}$. The number of SMEFT events is obtained at the upper bound of the corresponding Wilson coefficients from a single operator fit with 5\% syst. 
        }
        \label{fig:pThv1}
\end{figure}

Additionally, we show in Fig.~\ref{fig:pThv1} the number of events from each background process and the signal in each $p_T^h$ bin. In the case of the signal, we show the number of events for the SM case and the contribution from the $\mcO_{\varphi q}^{(3)}$, $\mcO_{uW,11}$, and $\mcO_{uW,22}$ operators with WCs at values representative of the computed bounds, $c_{\varphi q}^{(3)}= 3\times 10^{-3}$,  $c_{uW}^{11}=1.1\times 10^{-2}$ and $c_{uW}^{22}=3.3\times 10^{-2}$ with $\Lambda = 1~\tev$.

\section{Chiral Perturbation Theory and Meson Decays}\label{app:pidecay}

 Below the QCD confinement scale, the light mesons are regarded as the pNGBs from the spontaneous breaking of the chiral symmetry by the quark condensation $\langle q_L^\dagger q_R\rangle \neq 0$. In the non-linear $\sigma$ model described by Chiral Perturbation Theory ($\chi$PT), the pNGBs are packaged into
\begin{equation}
	U=\exp\left(\frac{i}{F_\pi}\pi^a\lambda^a\right)\;,
\end{equation}
where $\lambda^a,\; a=1\cdots 8$ are Gell-Mann matrices. Above the QCD scale, quarks are coupled to various spurions of the chiral symmetries, which provide additional building blocks for $\chi$PT~\cite{Pich:2018ltt}. At the two-quark level, one finds matrix-valued Lorentz scalar ($s,p$), vector ($v,a$) and tensor sources ($t$),
\begin{equation}
{\cal L}_{\rm QCD}=\bar q \left(\gamma^\mu[v_\mu+\gamma_5a_\mu]-[s-i\gamma_5p]+\sigma^{\mu\nu}t_{\mu\nu}\right)q \ .
\end{equation}
One also encounters the left/right equivalent of the vector/axial vector sources, $l_\mu \equiv v_\mu - a_\mu,r_\mu \equiv v_\mu + a_\mu$. Other spurions appear at the four-quark level and higher orders.

\subsection{$K\to \pi\pi$}\label{app:kpipi}
The WET operator $\bar{s}\gamma^\mu d q\gamma_\mu q$ will induce the non-leptonic $K\rightarrow\pi\pi$ decay~\cite{Buchalla:1995vs, Gilman:1979bc}. The decay rate can be computed via the chiral Lagrangian~\cite{Cirigliano:2003gt, Pich:2021yll}, 
\begin{equation}
    \mathcal{L}_{\rm weak}\supset -g_{8}G F_\pi^4\,\tr\left(P_{sd} L^\mu L_\mu\right)\;,
\end{equation}
where $L_\mu =i U^\dagger D_\mu U$ and the covariant derivative $D^\mu U=\partial^\mu U-i r^\mu U+i U l^\mu$, and $P_{sd}\equiv\frac12(\lambda_6-i\lambda_7)$ projects onto the $s\rightarrow d$ transition. Following Ref.~\cite{Pich:2021yll}, the chiral Lagrangian parameter $g_{8}$ is determined by the WET Wilson coefficients $C_i$ defined in Eq.~\eqref{WETlag}. For the operator ${\cal Q}_{q3}$ and its Wilson coefficient $C_{q3}$, to which the EW quark dipole SMEFT operator  $\mathcal{O}_{uW}$ contributes at the weak scale as shown in Eq.~\eqref{eq:C4F}, one finds,
\begin{equation}
    \Delta g_8=\frac{\sqrt2 \left(a_8^S-a_8^A\right)\Delta C_{q3} }{8V_{ud}V_{us}^*} \ ,
\end{equation}
 where $a_8^{S,A}$ are the numerical coefficients associated to the symmetric and antisymmetric combination of the chiral octets contained in $C_{q3}$ and can be computed with lattice methods. $\Delta C_{q3}$ is the contribution of $\mathcal{O}_{uW}$ which alters the parameter $g_8$. As shown in Fig.~\ref{fig:fcncgim}, the two-dipole insertion contributions to $C_{q3}$ are given by the same diagrams as for $C_{\ell 1}$, so that we find $\Delta C_{q3}=\Delta C_{\ell 1}$. The theoretical value of $g_8$, given by the Standard Model contribution to $C_{q3}$ and lattice simulations of $a_{8}^{S}=-0.20(24)$, $a_{8}^{A}=2.7(5)$, is $g_8^{\rm SM}=2.6(5)$~\cite{Pich:2021yll}. On the other hand, through measurements of the $K\rightarrow \pi\pi$ decay amplitude, we can determine its experimental value as $g_{8}^{\rm exp}=3.07(14)$. Assuming that $g_8$ falls within $1\sigma$ of the difference $\Delta g_8=g_8^{\rm SM}-g_8^{\rm exp}=0.47(52)$, one establishes the following bound on the W-dipole Wilson coefficients,
\begin{equation}
	|c_{uW}^{11}|\lesssim 13.2\left(\Lambda /\tev\right)^2\;.
\end{equation}
As explained around Eq.~\eqref{eq:WETops}, the EW quark dipole also contributes to the flavor-changing chromodipole~\cite{Hurth:2019ula},
\begin{equation}
    \mathcal{O}_{8g}^{sd(\prime)}=g_s m_{s}\left[\bar{s}P_{R(L)}\sigma_{\mu\nu}T^A d\right]G_A^{\mu\nu}\;,
\end{equation}
Its contribution to the $K\rightarrow\pi\pi$ hadronic matrix elements has been studied in Ref.~\cite{Buras:2018evv}, while its effect on the kaon direct CP violation is discussed in Ref.~\cite{Chen:2018vog}.
Systematically exploring the FCNC effects of $\mathcal{O}_{8g}^{sd}$ could be a promising avenue that we leave for future research.

\subsection{$\chi$PT with a Tensor Source}\label{app:tensorsource}
 
We discuss here the embedding in $\chi$PT of a tensor source which couples to quarks as follows,
\begin{equation}\label{eq:chiraltensor}
\mathcal{L}=\bar{q}_L\sigma^{\mu\nu}t^\dagger_{\mu\nu}q_R+\bar{q}_R\sigma^{\mu\nu}t_{\mu\nu}q_L\;,
\end{equation}
from which one reads its chiral transformation rules. In our case, where the tensor source arises from the SMEFT EW quark dipole, one finds at tree level,
\begin{equation}\label{eq:tmunu}
t_{\mu\nu}=\frac{c_{uW}v_h}{\Lambda^2}P_L^{\mu\nu\lambda\rho}W_{\lambda\rho}\;,\quad P_L^{\mu\nu\lambda\rho}=\frac14\left(g^{[\mu\lambda}g^{\nu]\rho}-i\varepsilon^{\mu\nu\lambda\rho}\right)\;.
\end{equation}
The projector $P_L$ appears due to the relation $\sigma^{\mu\nu}\gamma_5=\frac{i}{2}\epsilon^{\lambda\rho\mu\nu}\sigma_{\lambda\rho}$.
Following the notation in Ref.~\cite{Cata:2007ns}, it is also convenient to rewrite the chiral Lagrangian in terms of $\rho=\sqrt{U}$, and to trade the tensor source for the building blocks $t_\pm^{\mu\nu}$. Their definition, which can be extended to the field strengths $F_L,F_R$ of the vector sources $l,r$, is as follows, 
 \begin{equation}\label{eq: tfpm}
	t_{\pm}^{\mu\nu}=\rho^\dagger t^{\mu\nu}\rho^\dagger\pm\rho t^{\mu\nu\dagger}\rho\;,\quad f_\pm^{\mu\nu}=\rho F_L^{\mu\nu} \rho^\dagger \pm \rho^\dagger F_R^{\mu\nu} \rho \;,
\end{equation}
where  $F_{L(R)}^{\mu\nu}=P_{L(R)}^{\mu\nu\lambda\rho}F_{\lambda\rho}$. 
One then finds that the chiral invariant operators at the lowest order are given by~\cite{Cata:2007ns} 
\begin{equation}\label{eq:chptdipole}
	\mathcal{L}^{\chi {\rm PT}}\supset c_1F_0\langle t_+^{\mu\nu}f_{+\mu\nu}\rangle+c_2F_0^2\langle t_+^{\mu\nu}t_{+\mu\nu}\rangle + c_3F_0^2\langle t_{+\mu\nu}\rangle^2\;,
\end{equation}
where $\langle\cdots\rangle$ stands for the trace in the $u,d,s$ flavor space. As, in our case, $t_+^{\mu\nu}$ scales as $\Lambda^{-2}$, where $\Lambda$ is the SMEFT cutoff, the dominant contribution is given by the first term of Eq.~\eqref{eq:chptdipole}. 

\section{Other Effective Operators}
\label{app:Oeff}

As discussed in Sec.~\ref{sec:IHMFV}, in IHMFV, large left-handed FCNC may arise from the $LL$ type operators,
\begin{equation}\label{eq:LL}	\bar{Q}_L\widetilde{Y}_u\widetilde{Y}_u^\dagger Q_L\;.
\end{equation}
To check the size of these FCNC in  the model of Sec.~\ref{sec:simpmodel}, which realizes IHMFV, we use \texttt{Matchmakereft}~\cite{Carmona:2021xtq} to compute the dimension-6 $Q_LQ_L$-type operators induced by $X$ at 1-loop and the result is,
\bea\label{eq:LLWC}
	& \frac{16\pi^2}{\Lambda^2\tr[(M_X M_X^\dagger)^{-1}]}\left[c_{\varphi q}^{(1)}\right]_{ij}=\frac{g_1^2}{90}\left[g_1^2+10|\lambda_u|^2\left(1-\ln\frac{M_X^2}{\mu^2}\right)\right]\delta_{ij}\;,\quad \\
	&\frac{16\pi^2}{\Lambda^2 \tr[(M_X M_X^\dagger)^{-1}]} \left[c_{\varphi q}^{(3)}\right]_{ij}=\frac{g_2^2}{30}\left(9g_2^2-5|\lambda_u|^2\right)\delta_{ij}\;,\\
	&\frac{16\pi^2}{\Lambda^2\tr[(M_X M_X^\dagger)^{-1}]}\left[c_{qq}^{(1)}\right]_{ijk\ell}=-\frac{g_3^4}{30}\delta_{ij}\delta_{k\ell}+\frac{g_1^4}{540}\delta_{ij}\delta_{k\ell}\;,\\
	&\frac{16\pi^2}{\Lambda^2\tr[(M_X M_X^\dagger)^{-1}]}\left[c_{qq}^{(3)}\right]_{ijk\ell}=-\frac{g_3^4}{10}\delta_{i\ell}\delta_{kj}-\frac{3g_2^4}{20}\delta_{ij}\delta_{k\ell}\;,\\
	&\frac{16\pi^2}{\Lambda^2\tr[(M_X M_X^\dagger)^{-1}]}\left[c_{q\ell}^{(1)}\right]_{ijk\ell}=\frac{g_1^4}{90}\delta_{ij}\delta_{k\ell}\;,\quad \\
	&\frac{16\pi^2}{\Lambda^2\tr[(M_X M_X^\dagger)^{-1}]}\left[c_{q\ell}^{(3)}\right]_{ijk\ell}=-\frac{3g_2^4}{10}\delta_{ij}\delta_{k\ell}\;,\\
	&\frac{16\pi^2}{\Lambda^2\tr[(M_X M_X^\dagger)^{-1}]}\left[c_{qu}^{(1)}\right]_{ijk\ell}=-g_1^2\left[\frac{2g_1^2}{135}\delta_{k\ell}+\frac{|\lambda_u|^2}{648 }\left(17-18\ln\frac{M_X^2}{\mu^2}\right)\frac{(M_X M_X^\dagger)^{-1}_{kl}}{\tr[(M_X M_X^\dagger)^{-1}]}\right]\delta_{ij}\;,\\
	&\frac{16\pi^2}{\Lambda^2\tr[(M_X M_X^\dagger)^{-1}]}\left[c_{qu}^{(8)}\right]_{ijk\ell}=-g_3^2\left[\frac{4g_3^2}{5}\delta_{k\ell}+\frac{8|\lambda_u|^2}{9 }\frac{(M_X M_X^\dagger)^{-1}_{kl}}{\tr[(M_X M_X^\dagger)^{-1}]}
 \right]\delta_{ij}\;,\\
	&\frac{16\pi^2}{\Lambda^2\tr[(M_X M_X^\dagger)^{-1}]}\left[c_{qd}^{(1)}\right]_{ijk\ell}=\frac{g_1^4}{135}\delta_{ij}\delta_{k\ell}\;,\\
        &\frac{16\pi^2}{\Lambda^2\tr[(M_X M_X^\dagger)^{-1}]}\left[c_{qd}^{(8)}\right]_{ijk\ell}=-\frac{4g_3^4}{5}\delta_{ij}\delta_{k\ell}\;,\\
        &\frac{16\pi^2}{\Lambda^2\tr[(M_X M_X^\dagger)^{-1}]}\left[c_{qe}\right]_{ijk\ell}=\frac{g_1^4}{45}\delta_{ij}\delta_{k\ell}\;,
\eea
where we have used the Fierz identity
\begin{equation}
	\delta^{ij}\delta^{k\ell}\left(\bar{\psi}_i\gamma_\mu\psi_j\right)\left(\bar{\chi}_k\gamma^\mu\chi_\ell\right)=\delta^{ij}\delta^{k\ell}\left(\bar{\psi}_i\gamma_\mu\chi_\ell\right)\left(\bar{\chi}_k\gamma^\mu\psi_j\right)
\end{equation}
to simplify the expressions. We also neglected terms which arise from loops of $F$ and $S$, as they automatically conserve flavor. We nevertheless stress that such terms, suppressed by $m_S$ and $M_F$, are present.

\begin{figure*}[h!]
\centering
\begin{tikzpicture}[node distance=1.0cm and 0.75cm]
   \coordinate[vertex] (v1);
   \coordinate[label=left:$Q_L$, above left=of v1] (i1);
   \coordinate[label=left:$\bar{Q}_L$, below left=of v1] (i2);
   \coordinate[vertex, right=of v1] (v2);
   \node[] at ( $ (v1)!3!(v2) $ ) (v3) {};
   \node[] at ( $ (v1)!2!(v2) $ ) (vt) {};
   \coordinate[label=right:$H^\dagger$, above=of v3] (o1);
   \coordinate[label=right:$H$, below=of v3] (o2);
   \coordinate[vertex, above=0.7cm of vt] (v3);
   \coordinate[vertex, below=0.7cm of vt] (v4);
   \draw[fermion] (i1)--(v1);
   \draw[fermion] (v1)--(i2);
   \draw[photon] (v1)--node[above=1mm] {$V_\mu$}(v2);
   \draw[fermion] (v2)--(v3);
   \draw[fermion] (v3)--(v4);
   \draw[fermion] (v4)--node[below=1mm] {$X$}(v2);
   \draw[scalar] (o1)--(v3);
   \draw[scalar] (v4)--(o2);
  \end{tikzpicture}
\begin{tikzpicture}[node distance=1.0cm and 0.75cm]
   \coordinate[vertex] (v1);
   \coordinate[label=left:$Q_L$, above left=of v1] (i1);
   \coordinate[label=left:$\bar{Q}_L$, below left=of v1] (i2);
   \coordinate[vertex, right=of v1] (v2);
   \coordinate[vertex, right=of v2] (v3);
   \coordinate[vertex, right=of v3] (v4);
   \draw[fermion] (v3) arc[start angle=0, delta angle=180, radius=0.44cm];
   \draw[fermionnoarrow] (v3) arc[start angle=0, delta angle=-180, radius=0.44cm];
   \coordinate[label=right:$\bar{f}$, above right=of v4] (o1);
   \coordinate[label=right:$f$, below right=of v4] (o2);
   \node at (1.7,0.85) {$X$};
   \draw[fermion] (i1)--(v1);
   \draw[fermion] (v1)--(i2);
   \draw[photon] (v1)--node[above=1mm] {$V_\mu$}(v2);
   \draw[photon] (v3)--node[above=1mm] {$V_\mu$}(v4);
   \draw[fermion] (v4)--(o2);
   \draw[fermion] (o1)--(v4);
  \end{tikzpicture}
 \begin{tikzpicture}[node distance=1.0cm and 0.75cm]
   \coordinate[vertex] (v1);
   \coordinate[label=left:$Q_L$, above left=of v1] (i1);
   \coordinate[label=left:$\bar{Q}_L$, below left=of v1] (i2);
   \coordinate[vertex, right=of v1] (v2);
   \node[] at ( $ (v1)!3!(v2) $ ) (v3) {};
   \node[] at ( $ (v1)!2!(v2) $ ) (vt) {};
   \coordinate[label=right:$\bar{u}_R$, above=of v3] (o1);
   \coordinate[label=right:$u_R$, below=of v3] (o2);
   \coordinate[vertex, above=0.7cm of vt] (v3);
   \coordinate[vertex, below=0.7cm of vt] (v4);
   \draw[fermion] (i1)--(v1);
   \draw[fermion] (v1)--(i2);
   \draw[photon] (v1)--node[above=1mm] {$V_\mu$}(v2);
   \draw[fermion] (v2)--node[above=1mm] {$X$}(v3);
   \draw[scalar] (v3)--node[right=1mm] {$H$}(v4);
   \draw[fermion] (v4)--node[below=1mm] {$X$}(v2);
   \draw[fermion] (o1)--(v3);
   \draw[fermion] (v4)--(o2);
  \end{tikzpicture}

\caption{Feynman diagrams contributing to the one-loop matching between our UV model and the $LL$-type dimension-6 SMEFT operators, where $V_\mu = B_\mu,\, W_\mu$. These diagrams preserve the flavor symmetry.}
\label{fig:intX}
\end{figure*}
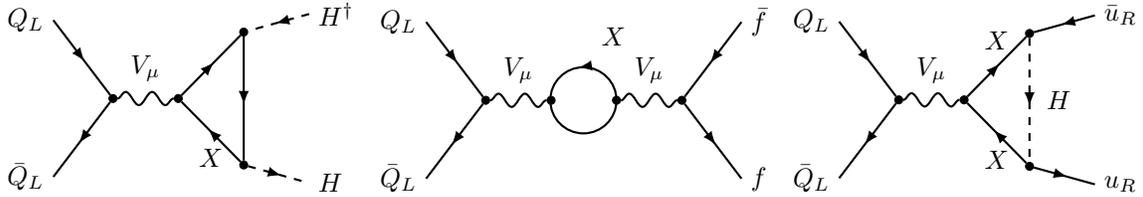

None of the $LL$-type operator in Eq.~\eqref{eq:LLWC} changes the quark flavor, because they are generated by the gauge interactions. 
This result can be understood from the diagrams relevant to the 1-loop matching after integrating out $X$, depicted in Fig.~\ref{fig:intX}.

We can further explain the observed pattern as follows. In the model of Sec.~\ref{sec:simpmodel}, the two-point function of the field $X$ is the only source of the inverse Yukawa coupling, $\langle X\bar{X}\rangle\simeq\widetilde{Y}_u/M$. At one loop, one finds that large FCNC mediated by the operators in Eq.~\eqref{eq:LL} can only be generated through the coupling between $\bar{Q}_L$, $X$ and another operator $\mathcal{O}$. The mass dimension of $\mathcal{O}$ in the renormalizable theory is 1 and we collectively denote the properties of $\mathcal{O}$ as,
\begin{equation}\label{eq:flvsup}
	\mathcal{L}\supset\widetilde{\lambda}\bar{Q}_L X \mathcal{O}\;, \quad [\mathcal{O}]=1\;.
\end{equation}
The operator $\mathcal{O}$ can not be the mass mixing which is removable by the field redefinition. The only two candidates are the singlet scalar $S$ and the gauge boson $V_\mu$ which appeared in the covariant derivative $D=\partial+ig V$,
\begin{equation}
\mathcal{L}\supset\widetilde{\lambda}_V\bar{Q}_L\slashed{D} X+\widetilde{\lambda}_S\bar{Q}_LS X\;.
\end{equation}
The first term does not mix quark flavors, because both the kinetic mixing and the mass mixing can be simultaneously eliminated through the rescaling and the rotation transformation. In the simple model of Eq.~\eqref{eq:simpUV}, the $U(1)_M$ symmetry forbids the $Q_LSX$ coupling. Then, there is no available field that can play the role of $\mathcal{O}$. In general, as long as interactions of the type in Eq.~\eqref{eq:flvsup} are absent in the renormalizable UV model, the FCNC operators Eq.~\eqref{eq:LL} will be suppressed in the IR EFT. We checked this in our model by computing all the dimension-6 effective operators and found that there are no $LL$-type flavor-violating operators.

Being flavor-universal and loop-suppressed, those operators accompanying the dipole are very weakly bounded by low-energy flavor probes. However, they can source new collider signals beyond those studied in Section~\ref{sec:EWdipole_hadron_coll}, which would constitute complementary tests of this specific model. In the first sections of this paper, we did not commit to a specific UV scenario, hence we focused on the phenomenology associated with the dipole. Nevertheless, it is worth assessing how this phenomenology is complemented in the case of the UV model which we analyzed, and when it is representative of the complete physics. At the level of Wh production, which was the focus of Section~\ref{sec:EWdipole_hadron_coll}, $c_{\varphi q}^{(3)}$ and $c_{uW}^{11}$ contribute equally: the slightly stronger bounds on $c_{\varphi q}^{(3)}$ from Table~\ref{tab:bounds_summary} are offset by the fact that this WC is naturally a factor of a few smaller than $c_{uW}^{11}$ in our model. Thus, in this channel, the dipole phenomenology is representative of the physics and of the FCC sensitivity to the whole model. 4-Fermi operators that are generated by the model are constrained by the Drell-Yan process, leading at the LHC to a bound on their WCs of the order of $10^{-3}$ TeV${}^{-2}$. This happens to be a factor of a few above the natural value that they take in our model for BSM masses around a TeV. Evaluating how the Drell-Yan bounds will change at the FCC-hh is certainly interesting, and has not been done to the best of our knowledge, but it lies beyond the scope of this work and we leave it for a future publication. Generally, a multi-channel analysis of the model will likely strengthen the sensitivity at the FCC-hh, and would be necessary in order to pinpoint the precise model, were a signal found in a specific channel.

\bibliography{dipolediboson} 
\end{document}